\documentclass[letterpaper]{JHEP3}
\usepackage{epsfig,amsmath}

\usepackage{subfig}

\newcommand{\pd}[2]{\frac{\partial #1}{\partial #2}}

\title{A numerical approach to finding general stationary vacuum black holes}

\author{
Alexander Adam, Sam Kitchen and Toby Wiseman \\ Blackett Laboratory, Imperial College London, London SW7 2AZ, UK \\\email{alexander.adam06 , sam.kitchen03, t.wiseman@imperial.ac.uk}
}

\date{June 2011}

\abstract{

The Harmonic Einstein equation is the vacuum Einstein equation supplemented by a gauge fixing term which we take to be that of DeTurck. 
For static black holes analytically continued to Riemannian manifolds without boundary at the horizon this equation has previously been shown to be elliptic, and Ricci flow and Newton's method provide good numerical algorithms to solve it. 
Here we extend these techniques to the arbitrary cohomogeneity stationary case which must be treated in Lorentzian signature.
For stationary spacetimes with globally timelike Killing vector the  Harmonic Einstein  equation is elliptic. In the presence of horizons and ergo-regions it is less obviously so.
Motivated by the Rigidity theorem we study a class of stationary black hole spacetimes, considered previously by Harmark, general enough to include the asymptotically flat case in higher dimensions.
We argue the Harmonic Einstein equation consistently truncates to this class of spacetimes giving an elliptic problem. The Killing horizons and axes of rotational symmetry are boundaries for this problem and we determine boundary conditions there.
As a simple example we numerically construct 4D rotating black holes in a cavity using Anderson's boundary conditions.
We demonstrate both Newton's method and Ricci flow to find these Lorentzian solutions.

}

\begin{document}

\section{Introduction}

Stationary vacuum black holes in spacetime dimension $D > 4$ have received much attention over the past decade, and many remarkable results have emerged \cite{Kol:2004ww,Harmark:2007md,Emparan:2008eg}. In asymptotically flat $D=4$ spacetime Kerr is unique. In $D>4$ beyond the Myers-Perry solution \cite{MyersPerry} which generalizes Kerr, the space of solutions thought to exist is very exotic \cite{Emparan:2007wm} and  there may yet be further solutions discovered.
String theory famously requires more than 4 dimensions for consistency, and hence provides a powerful physical motivation to think about these exotic solutions.
Perhaps the strongest motivation to consider higher dimensions and also exotic boundary conditions is the $AdS$-$CFT$ correspondence, which relates many of these interesting solutions and other gravitational phenomena to physical phenomena found in certain strongly coupled conformal field theories and gauge theories where direct field theory calculations are extremely hard.
The majority of the solutions conjectured to exist are unlikely to be found analytically using current techniques.
The problem of solving the Einstein equations to find them reduces to solving partial differential equations for the components of the metric. Since these exotic solutions are of cohomogeneity greater than one it is not clear this can be done analytically in general.
Thus we are forced to consider numerical methods to construct such solutions. 
The aim of this paper is to find numerical methods that are geometrically elegant, apply for arbitrary cohomogeneity and topology, and that function well in practice.

From the classic 4$D$ stationary uniqueness theorems (see for example \cite{Carter}) we understand that the stationary problem of finding vacuum black holes should be thought of as an elliptic PDE system which is treated as a boundary value problem, where asymptotic conditions are imposed together with regularity at the horizon. Rigidity implies the existence of axisymmetry, and then the problem reduces to one depending non-trivially on two coordinates (`cohomogeneity two'). Taking the Weyl-Papapetrou form for the metric then manifests the PDE problem as an elliptic one.
Black hole uniqueness theorems have been extended to $D>4$ in the stationary case of $D-2$ commuting Killing directions \cite{Morisawa:2004tc,Hollands:2007aj} where by reducing on these Killing directions the problem can be viewed as a 2$D$ elliptic one on the orbit space of these isometries. 
For non-trivial dependence on only two dimensions the stationary numerical problem has been extensively studied in the context of relativistic stars for decades \cite{W72, BS74, BI76, Stergioulas}, using formulations based on the Weyl-Papapetrou form of the metric where a sufficient subset of the Einstein equations are elliptic. These methods have been applied to exotic static and stationary 4$D$ black holes \cite{Kleihaus:1997ic,Kleihaus:2000kg}. 
In \cite{star,Wiseman:2002zc} it was shown how to apply cohomogeneity two methods to higher dimensions, without having $(D-2)$ commuting Killing vectors. A form for the metric analogous to the Weyl ansatz is taken and the Einstein equations reduce to a set of elliptic ones for the metric components, together with constraints which may be satisfied provided appropriate boundary conditions are imposed.

The main challenge to formulating the problem for general cohomogeneity is to present the Einstein equations as an elliptic system in an elegant and covariant manner. In this paper we build on the previous algorithms proposed in \cite{KitchenHeadrickTW} to find static vacuum black holes.
 In this method a static solution is analytically continued to a Riemannian manifold. If we are considering a black hole, then provided it has a single non-extremal horizon or multiple horizons sharing the same (non-zero) surface gravity, then the periodicity of Euclidean time can be chosen such that this Euclidean signature manifold has no boundary at the location of the horizon and is perfectly smooth there. The only boundaries in the problem arise from the region far from the horizon where one may consider various boundary or asymptotic conditions \cite{PFLuciettiTW}. An elegant feature of this approach is that static vacuum Lorentzian geometries can then be treated simply as a particular case of  the problem of finding Riemannian Einstein metrics which is more general and has other interesting applications \cite{CY,dP3}. 
Instead of solving the Einstein equation directly, one solves the Harmonic Einstein equation with DeTurck's choice of gauge fixing term.\footnote{
Note that in the previous work \cite{KitchenHeadrickTW,PFLuciettiTW} the `Harmonic Einstein' equation with DeTurck gauge fixing was referred to as the `Einstein-DeTurck' equation. We will use the  `Harmonic Einstein'  terminology here to avoid potential confusion with the `Ricci-DeTurck' flow.
} 
For  Euclidean signature this is elliptic and can be treated as a boundary value problem.  
The Ricci-DeTurck flow or Newton method provide explicit algorithms to find solutions \cite{KitchenHeadrickTW}. Numerical Ricci flows have been considered in \cite{Garfinkle:2003an, HeadrickTW, Holzegel:2007zz, dP3, Headrick:2007fk, KitchenHeadrickTW}.

This approach is closely related to the use of generalized harmonic coordinates in the hyperbolic context \cite{Friedrich, Garfinkle:2001ni}. One solves the Harmonic Einstein equation to yield a solution to the Einstein equation in generalized harmonic gauge. There is an important and subtle difference between the hyperbolic and elliptic contexts.
For dynamics the vanishing of this gauge fixing term is imposed on the initial data, and the contracted Bianchi identity implies it remains zero at later times. Whilst one solves the Harmonic Einstein equation, one only considers this equation in the neighbourhood of a solution to the Einstein equation.
In contrast in the Riemannian elliptic case one does not have this luxury. One must impose physical boundary conditions and then solve the Harmonic Einstein equation subject to these. Since one does not know the solutions which correspond to solutions of the Einstein equation, the whole problem being to find these, one cannot consider the Harmonic Einstein equation only in the neighbourhood of such Einstein solutions, but must consider it generally subject to the boundary data. In principle there may exist solutions to the Harmonic Einstein equation, termed Ricci solitons, where the solution is not Einstein, and the DeTurck gauge fixing term does not vanish. 
In practice one can check whether the solution found is a soliton or not. However, the existence of solitons is rather constrained. Bourguignon \cite{Bourguignon} has proven on a compact manifold without boundary such solitons cannot exist, and in \cite{PFLuciettiTW} it was argued using a simple maximum principle that in this static case with various boundary and asymptotic conditions no solitons exist. 

Our aim here is to develop general methods to find stationary black holes. However this Riemannian approach for static solutions cannot be applied to the stationary case since there is no analytic continuation of the stationary vacuum Einstein equations. The key observation is that the Harmonic Einstein equation of the Riemannian signature black hole gives precisely the same equations for the metric components if one continues back to Lorentzian signature provided one uses coordinates adapted to the static symmetry. Whilst in general the Lorentzian Harmonic Einstein equation has hyperbolic signature, this shows that when one considers the Harmonic Einstein equation restricted to static Lorenztian metrics it is elliptic and can be solved as a boundary value problem. The only difference is that now the Lorentzian horizon represents a physical boundary - in the Riemannian continuation the static adapted coordinate chart has a boundary at the horizon, in analogy to the origin of polar coordinates. One may determine the boundary conditions in these static coordinates by considering smoothness in a chart that does cover  the horizon (analogous to Cartesian coordinates).
The key result of this paper is to show that the Harmonic Einstein equation restricted to a class of Lorentzian stationary metrics motivated by the Rigidity theorem \cite{Hollands:2006rj,Moncrief:2008mr}, general enough to consider wide classes of stationary black holes which include the asymptotically flat case,  is elliptic. The horizon must be thought of as a boundary, and there are well posed boundary conditions that can be given there. The resulting elliptic system can be solved by Ricci flow on this space of Lorentzian stationary metrics, or by Newton's method.

For simplicity we restrict our discussion here to vacuum solutions with no cosmological term and only non-extremal Killing horizons.
The structure of the paper is then as follows. We begin in section \S\ref{sec:method} by reviewing the approach to finding static solutions using the Euclidean continuation and Harmonic Einstein equation. Then in section \S\ref{sec:static} we show how to reinterpret this from an entirely Lorentzian perspective, demonstrating the equation is elliptic and that regularity conditions can be given at the horizon, which in Lorentzian signature appears as a boundary. In
\S\ref{sec:stationary} we consider the character of the Lorentzian stationary Harmonic Einstein equation. We show that for globally timelike stationary Killing vector (i.e. stationary spacetimes without horizon or ergo-regions) this is again elliptic. From this analysis it is clear that the threat to ellipticity in the stationary case is due to ergo-regions.
In \S\ref{sec:blackholes} we make an ansatz for stationary black hole spacetimes motivated by the Rigidity theorem, general enough to include the asymptotically flat case. This is the ansatz that Harmark  \cite{Harmark:2009dh}  has used to consider classification of higher dimensional black holes. 
Following from Rigidity, Killing symmetry in the directions of angular or linear motion of the horizon is assumed, and this is crucial for maintaining ellipticity in the presence of ergo-regions. 
Based on the analytic treatment of the stationary problem for the uniqueness theorems, we assume the orbit space is a smooth Riemannian manifold, and the horizon and any axes of symmetries of the angular motion are boundaries of this space.
We are then able to conclude that the Harmonic Einstein equation is elliptic when restricted to such stationary spacetimes.
We determine the boundary conditions at the horizon and rotational symmetry axes in an analogous manner to those in the Lorentzian static case, and show they are compatible with finding non-soliton solutions to the Harmonic Einstein equations ie. solutions of the stationary Einstein equations.
We conclude the paper in section \ref{sec:example} by giving a simple numerical example of these techniques. We numerically find 4$D$ rotating black holes in a cavity. We use Anderson's boundary condition, taking the boundary to be conformal to the product of time and a round sphere with fixed trace of its extrinsic curvature \cite{Anderson1}. We demonstrate the use of both Newton's method and Ricci flow as algorithms to find the solutions.

\section{Review of the Harmonic Einstein equation, Ricci flow and Newton's method for static geometries}
\label{sec:method}

We now review in more detail the framework for finding static vacuum black holes that we wish to extend to the stationary case.
Since we will consider vacuum solutions with no cosmological term we are interested in finding Ricci flat static metrics.
We analytically continue time to give a Euclidean metric. We take all the Killing horizons to have the same surface gravity so that we may choose Euclidean time to have the appropriate period such that we have a smooth Riemannian manifold, and no boundaries associated to the horizons. Thus the problem is now to find Ricci flat Riemannian metrics with a $U(1)$ isometry which is generated by a hypersurface orthogonal Killing vector, and which has fixed points of the isometry at the locations which continue to Killing horizons.

Following our previous work, instead of solving the Ricci flatness condition $R_{\mu\nu} = 0$, we instead introduce the tensor, 
\begin{eqnarray}
{R}^H_{\mu\nu} \equiv R_{\mu\nu} - \nabla_{(\mu} \xi_{\nu)} 
\end{eqnarray}
which we will refer to as the Ricci-DeTurck tensor,
where the vector field $\xi^\mu$ is constructed as,
\begin{eqnarray}
\xi^\mu = g^{\alpha\beta} \left( \Gamma_{\alpha\beta}^{~~\mu} - \bar{\Gamma}_{\alpha\beta}^{~~\mu} \right)
\end{eqnarray}
from a fixed smooth reference connection $\bar{\Gamma}_{\alpha\beta}^{~~\mu}$ on the manifold. The term $\nabla_{(\mu} \xi_{\nu)}$ is the `DeTurck term' introduced by DeTurck in the context of Riemannian geometry and later used by him to show Ricci flow is parabolic \cite{DeTurck}. We note that the vector field $\xi$ is globally defined, being the difference of our Levi-Civita and reference connection. This term is a global analog of the local term considered in generalized harmonic gauge fixing \cite{Friedrich,Garfinkle:2001ni} which is now extensively used in numerical simulation following Garfinkle's work.

We focus on solving the `Harmonic Einstein equation' $R^H_{\mu\nu} = 0$. The Ricci-DeTurck tensor has the property that it exhibits a simple structure in its second derivative terms, so that, 
\begin{eqnarray}
{R}^H_{\mu\nu} \equiv R_{\mu\nu} - \nabla_{(\mu} \xi_{\nu)} \sim -\frac{1}{2} g^{\alpha\beta} \partial_\alpha \partial_\beta {g}_{\mu\nu} + \ldots
\end{eqnarray}
where $\ldots$ refer to terms of lower order in derivatives. Then for a Riemannian signature manifold the Harmonic Einstein equation is clearly elliptic, and can be solved as an elliptic boundary value problem in the metric components $g_{\mu\nu}$. Whilst the Ricci tensor $R_{\mu\nu}$ shares the static symmetry of the metric $g_{\mu\nu}$, $R^H_{\mu\nu}$ does not unless we choose the reference connection appropriately. We do so by taking the reference connection to be the Levi-Civita connection of a smooth reference metric, $\bar{g}_{\mu\nu}$, so that,
\begin{eqnarray}
\xi_\alpha = g^{\mu\nu} \left( \bar{\nabla}_\mu g_{\nu \alpha} -  \frac{1}{2} \bar{\nabla}_\alpha g_{\mu\nu} \right)
\end{eqnarray}
and we take this reference metric to share the static isometry. Here, $\bar{\nabla}$ is the covariant derivative with respect to the metric $\bar{g}_{\mu\nu}$. This ensures $R^H_{\mu\nu}$ is symmetric with respect to the static isometry. Both metric and reference metric continue to Lorentzian spacetimes with the same static Killing horizons with equal surface gravities with respect to Killing time.

The aim is to solve the Harmonic Einstein equation to yield a Ricci flat solution with $\xi^\mu = 0$ which specifies that the metric is presented in the generalized harmonic gauge defined by our reference metric. However in general there will exist solutions which are not Ricci flat, but solve $R_{\mu\nu} = \nabla_{(\mu} \xi_{\nu)}$ so that $\xi^\mu$ does not vanish. Such solutions are called Ricci solitons. Fortunately the existence of Ricci solitons is quite constrained as we shall now discuss.
The contracted Bianchi identity for the Harmonic Einstein equation implies $\xi$ obeys the elliptic linear equation,
\begin{eqnarray}
\label{eq:bianchi}
\mathcal{O}_\mu^{~\nu} \xi_\nu \equiv \nabla^2 \xi_\mu + R_{\mu}^{~\nu} \xi_\nu = 0 \, .
\end{eqnarray}
The boundary conditions for the metric may be taken as giving boundary conditions for the vector field $\xi$. Let us consider a vector field $v^\mu$ with the same boundary behaviour as that of $\xi^\mu$. We expect that our boundary conditions are such that the problem,
\begin{eqnarray}
\label{eq:Ov}
\mathcal{O}_\mu^{~\nu} v_{\nu} = 0
\end{eqnarray}
is a well posed elliptic linear PDE in the vector components $v^\mu$, and that $v^\mu = 0$ is a solution. For example, taking asymptotically flat boundary conditions then $\xi^\mu \to 0$ in the asymptotic region \cite{PFLuciettiTW}. Alternatively taking Anderson's boundary conditions where the conformal class of the metric is fixed on a boundary, together with the trace of the extrinsic curvature \cite{Anderson1, PFLuciettiTW} then one imposes $\xi^\mu = 0$. For the boundary conditions relevant for a vacuum Randall-Sundrum brane \cite{RSI,RSII}, the normal component of $\xi^\mu$ to the brane boundary vanishes, and the tangential components of $\xi^\mu$ have Neumann boundary conditions. In all these three examples the associated problem has elliptic boundary conditions for the vector $v^\mu$. In the first two these ensure $v^\mu$ vanishes, whereas in the latter one, they are consistent with $v^\mu$ vanishing but do not impose it.

A necessary condition for a Ricci soliton to exist is that the linear elliptic operator $\mathcal{O}$ must have a non-trivial kernel. For given boundary conditions on the metric we may view this as an obstruction to the existence of Ricci solitons. A well known result is that for compact manifolds without boundary there are no Ricci soliton solutions for any choice of vector field. Using a simple maximum principle based on the contracted Bianchi identity one can show that with a variety of boundary conditions, including asymptotically flat, Kaluza-Klein and Anderson's boundary conditions, then again no Ricci solitons can exist \cite{PFLuciettiTW}.

However, even assuming Ricci solitons do exist, then the well-posedness of the Harmonic Einstein equation implies that generically\footnote{There may be special points in moduli space of Ricci flat solutions where a normalizable zero mode of the linearisation of the Harmonic Einstein equation exists, and this lifts to the non-linear equation to generate a deformation to a soliton solution. If a branch of solitons meets a branch of Ricci flat solutions this would be the case.} solutions should be locally unique. Hence a Ricci flat solution cannot be perturbatively close to a soliton solution. Therefore even if solitons exist, 
it is straightforward in principle to identify whether a solution that has been found is a soliton or not, the most obvious way being to compute $\xi$ and determine if it vanishes. 

Here we will now briefly review Anderson's boundary conditions since we will use them later in our numerical example. Anderson has shown the surprising result that one cannot impose the most obvious Dirichlet boundary condition, namely fixing the induced metric on a boundary, since it does not give rise to a regular elliptic system. 
Instead he has argued that one can fix the conformal class of the induced metric on a boundary together with the trace of the extrinsic curvature \cite{Anderson1}.
The metric in the vicinity of such a boundary can be written as,
\begin{eqnarray}
ds^2 = N dr^2 + N_a dr d {x}^a + {g}_{ab} d{x}^a d{x}^b
\end{eqnarray}
where the boundary is the hypersurface $r = 0$ and the remaining tangential coordinates are ${x}^a$ for $a = 1, \ldots, D-1$. 
For a static non-extremal black hole continued to Euclidean signature, the boundary will be a product of the Euclidean time circle with some spatial geometry. Fixing the conformal class of the induced metric and trace of the extrinsic curvature locally gives $D(D-1)/2$ conditions on the metric functions. However since the metric has $D(D+1)/2$ independent components we require another $D$ conditions. Recall our previous discussion, that on any boundaries or asymptotic regions we must ensure that $\xi$ has a behaviour such that the associated linear problem in equation \eqref{eq:Ov} defined above is well posed and has trivial solution $v^\mu = 0$. In order to achieve this the remaining $D$ conditions are fixed by imposing $\xi^\mu = 0$ on the boundary.

For Riemannian metrics we may solve the Harmonic Einstein equation using two distinct methods. Firstly we may use the DeTurck flow,
\begin{eqnarray}
\frac{d}{d\lambda} g_{\mu\nu} = -2 R^H_{\mu\nu}
\end{eqnarray}
beginning with an initial guess metric, and then flowing in the auxiliary time $\lambda$, hoping to end at a fixed point of the flow, $R^H_{\mu\nu} = 0$. The flow is strongly parabolic provided $R^H_{\mu\nu}$ is elliptic. Furthermore for our choice of reference metric it preserves the static isometry of $g$.
A very elegant feature of this flow is that it is diffeomorphic to the Ricci flow $\frac{d}{d\lambda} g_{\mu\nu} = -2 R_{\mu\nu}$, since the DeTurck term $\nabla_{(\mu} \xi_{\nu)}$ just introduces an infinitessimal diffeomorphism at each point along the flow. Hence while the flow in the space of metrics will depend explicitly on the choice of reference connection $\bar{\Gamma}$, geometrically the flows do not depend on this choice. An important consequence is that the basin of attraction of a fixed point of the flow is geometrically invariant, and doesn't depend on the choice of $\bar{\Gamma}$. 
One important drawback of this method is that, as discussed in \cite{HeadrickTW} vacuum black holes of interest are often unstable fixed points of Ricci flow, due to the existence of Euclidean negative modes of the Lichnerowicz operator \cite{GPY}. However, the Ricci-DeTurck flow may still be used as an algorithm. If the fixed point of interest has $n$ negative modes, then as described in \cite{HeadrickTW}, an $n$ parameter family of initial guesses must be taken, and the $n$ parameters tuned in the sense of a shooting problem in order to reach the fixed point of interest. 

The second method is to consider solving the equations $R^H_{\mu\nu}$ in a non-local way, using the canonical method to solve non-linear equations, the Newton's method. Writing the linearisation of $R^H$ as,
\begin{eqnarray}
R^H_{\mu\nu}[ g + \epsilon \delta g](x) = R^H_{\mu\nu}[g](x) + \epsilon {\Delta}[g](x)_{\mu\nu}^{~~~\alpha\beta} \delta g_{\alpha\beta}(x) + O(\epsilon^2)
\end{eqnarray}
we may think of this as an (infinite dimensional) vector equation, with $O(g)$ a matrix, with indices constructed from the spacetime index pair $\mu\nu$ and the point $x$ on the manifold. In practice one will introduce a truncation of the spacetime, for example using finite difference or spectral methods. Let us denote the finite dimensional collective index which represents the spacetime and position/mode using $A,B$, so that we may write $g_{\mu\nu}(x)$ as the vector $g_A$. Then we write the above equation as,
\begin{eqnarray}
R^H_A( g + \epsilon \delta g) = R^H_A (g) + \epsilon {\Delta}(g)_{A}^{~B} \delta g_{B} + O(\epsilon^2)
\end{eqnarray}
The Newton method iteratively improves a guess metric $g^{(i)}_A$ as,
\begin{eqnarray}
g^{(i+1)}_A - g^{(i)}_A = - ({\Delta}(g^{(i)})^{-1})_A^{~B} R^H_{B}(g^{(i)})
\end{eqnarray}
As for the Ricci flow method, the Newton method preserves the static isometry of the metric.
This method has the important advantage over the Ricci-DeTurck flow method that it is not sensitive to negative modes of the Lichnerowicz operator. However it does assume that the linear problem ${\Delta} \cdot v = R^H$ can be solved for $v$. In practice robust methods exist to solve such (finite dimensional) linear systems, such as biconjugate gradient, which are insensitive to the spectrum of ${\Delta}$, provided there are no zero modes. Thus a single initial guess will suffice, rather than having to tune a family of initial guesses.
An important disadvantage of the Newton method over the Ricci-DeTurck flow is that it is not geometric in the sense that the path taken by the algorithm in the space of geometries will depend explicitly on the choice of reference connection $\bar{\Gamma}$. This implies that the basin of attraction of a solution will also depend on this choice.

\section{Static spacetimes from a Lorentzian perspective}
\label{sec:static}

Instead of immediately now considering stationary spacetimes, it is instructive to first consider static spacetimes from a Lorentzian perspective. The  Harmonic Einstein equation is not elliptic for a general Lorentzian manifold, and without ellipticity one would not expect to be able to impose the various boundary conditions that we require physically in a well posed manner. However, consider a chart away from any horizon which manifests the static symmetry, 
\begin{eqnarray}
\label{eq:staticLor}
ds^2 = g_{\mu\nu} dx^\mu dx^\nu =  - N(x) dt^2 + h_{ij}(x) dx^i dx^j
\end{eqnarray}
so that $N > 0$. We may regard Euclidean time as being fibered over a base, the Riemannian manifold which we shall denote $\mathcal{M}$ with Euclidean metric $ds^2_{\mathcal{M}} = h_{ij}(x) dx^i dx^j$.
Furthermore with the choice  that our reference metric is also static with respect to $\partial / \partial t$, so that,
\begin{eqnarray}
\bar{ds}^2 = \bar{g}_{\mu\nu} dx^\mu dx^\nu=  - \bar{N}(x) dt^2 + \bar{h}_{ij}(x) dx^i dx^j
\end{eqnarray}
again with $\bar{N} > 0$ and $\bar{h}_{ij}$ a smooth Euclidean metric, then $R^H_{\mu\nu}$ shares the static symmetry. Due to this static symmetry the Harmonic Einstein equation $R^H_{\mu\nu} = 0$ thought of as PDEs for the metric components of $g$ is invariant under an analytic continuation $t \rightarrow  \tau = i t$. Hence we immediately see that the Harmonic Einstein equation restricted to Lorentzian static metrics and reference metrics is elliptic.
Furthermore, Ricci-DeTurck flow yields precisely the same flow equations for the metric functions $N$ and $h_{ij}$ above in either Lorentzian signature, or under continued Euclidean signature.

Explicitly the static Ricci-DeTurck tensor has components,
\begin{eqnarray}
R^H_{tt} &=&  - \frac{1}{2} \hat{\nabla}^{i}( \partial_{i} N)  + \frac{1}{2 N} ( \partial^{i} N)( \partial_{i} N) -  \frac{1}{2} \hat{\xi}^{k} \partial_{k} N  - \frac{1}{4 N} \bar{h}_{(-1)}^{mi} ( \partial_{m} \bar{N})( \partial_{i} N)   \nonumber \\
R^H_{ti}& = & 0 \nonumber \\
R^H_{ij} &=&  \hat{R}_{ij} - \hat{ \nabla}_{(i} \hat{\xi}_{j)} - \frac{1}{4 N^2} ( \partial_{i} N)( \partial_{j} N) - \frac{1}{2} h_{k(i}  \hat{ \nabla}_{j)}  \left( \frac{1}{N} \bar{h}_{(-1)}^{km} \partial_{m} \bar{N} \right)
 \end{eqnarray}
  where indices are contracted and covariant derivatives $\hat{\nabla}$ are with respect to the base metric $h_{ij}$. To avoid confusion we use the notation $\bar{h}_{(-1)}^{ij}$ for the inverse metric to $\bar{h}_{ij}$, so that $\bar{h}_{ik} \, \bar{h}_{(-1)}^{kj}  = \bar{h}^{(-1)ij} \bar{h}_{jk} = \delta^{i}_{k}$. The vector field $\hat{\xi}^i$ is the DeTurck vector of the base metric, namely,
\begin{eqnarray}
\label{eq:baseDeTurck}
\hat{\xi}^i = h^{jk} \left( \hat{\Gamma}^i_{~ jk} - \bar{\hat{\Gamma}}^i_{~ jk} \right)
\end{eqnarray}
where $\hat{\Gamma}^i_{~ jk} $ is the connection for $h_{ij}$ and $\bar{\hat{\Gamma}}^i_{~ jk} $ is the connection for $\bar{h}_{ij}$. We see the terms that contain two derivatives acting on the metric components of $g$ are of elliptic form provided $h_{ij}$ is a Riemannian metric. We also explicitly see that $R^H_{\mu\nu}$ is symmetric under the static isometry.

In the exterior of any horizons we have $N, \bar{N} > 0$. However now consider a non-extremal Killing horizon, where $\partial / \partial t$ has fixed action. Now $N$ and $\bar{N}$ vanish at the horizon. The base $h_{ij}$ must remain a smooth Riemannian geometry where $N = 0$ for the spatial horizon to have a well defined geometry, and we choose the same for $\bar{h}_{ij}$.
 In the Riemannian case we know that if we have have a single non-extremal horizon, or multiple horizons with the same surface gravity, then we may make Euclidean time periodic as, $\tau \sim \tau + 2 \pi / \kappa$ for some appropriate choice of constant $\kappa$, such that there is no boundary at the horizon(s) and the geometry is smooth there. Furthermore we know that $R^H_{\mu\nu}$ is a smooth tensor on this geometry, and hence the Ricci-DeTurck flow and Newton methods preserve the smoothness and lack of boundary at the Riemannian horizon.
However, in the chart above with $t \rightarrow \tau = i t$, the metric becomes,
\begin{eqnarray}
\label{eq:staticEuc}
ds^2 =  + N(x) d\tau^2 + h_{ij}(x) dx^i dx^j
\end{eqnarray}
and the chart does not cover the horizon where $N = 0$. Such coordinates adapted to the static symmetry are analogous to polar coordinates, and fail at the polar origin, the horizon. To manifest the smoothness of the Riemannian manifold one must go to `Cartesian' coordinates. Taking coordinates in the base adapted to the horizon such that $x^i = (r , x^a)$ where $r =0 $ is the horizon, we write,
\begin{eqnarray}
\label{eq:staticpolarEuc}
ds^2 =  + r^2 V d\tau^2 + U \left( dr + r \, U_a dx^a \right)^2 + h_{ab} dx^a dx^b
\end{eqnarray}
where the metric functions are functions of $r$ and $x^a$. Changing to coordinates,
\begin{eqnarray}
a = r \sin{\kappa \tau} \, , \quad b = r \cos{\kappa \tau} 
\end{eqnarray}
provides a good chart covering the horizon, such that the metric components are smooth functions, provided that
$V, U, U_a, h_{ab}$ are smooth ($C^\infty$) functions of $r^2$ and $x^a$, and,
\begin{eqnarray}
V = \kappa^2  U
\end{eqnarray}
at the horizon $r =0$ \cite{KitchenHeadrickTW, PFLuciettiTW}. Precisely the same conditions will apply to the reference metric which is also required to be smooth with no boundary at $r = 0$.
We now see that instead of using a good chart which does not manifest the static isometry, we might just as well use the original `polar' chart \eqref{eq:staticpolarEuc} and simply treat the horizon as a boundary, and determine the boundary behaviour using the regular chart, namely that $V, U, U_a, h_{ab}$ are smooth in $r^2$ and $x^a$, and that $V = \kappa ^2 U$ at $r = 0$ where the constant $\kappa$ determines the angular period of Euclidean time, and hence the temperature of the solution. In practice, adapting coordinates to the static symmetry, and indeed any other isometries, is important numerically to simplify the problem, and this is precisely the approach taken in previous work \cite{HeadrickTW,KitchenHeadrickTW}. As noted above taking smooth coordinates we know $R^H_{\mu\nu}$ preserves smoothness and lack of boundary at the horizon. Whilst in adapted coordinates we have a boundary at the horizon the same must be true, namely that the smoothness and regularity conditions above apply equally well to the tensor $R^H_{\mu\nu}$ which must be a smooth tensor on $g$ and we may explicitly check that,
\begin{eqnarray}
R^H =  + r^2 f d\tau^2 + g \left( dr + r \, g_a dx^a \right)^2 + r_{ab} dx^a dx^b
\end{eqnarray}
where the functions $f, g, g_a$ and $r_{ab}$ are smooth in $r^2, x^a$, and in addition $f = \kappa^2 g$ at $r = 0$.  However, the simpler way to see that this must be the case is to remember that in the smooth Cartesian coordinates $(a,b, x^a)$ then $g_{\mu\nu}$ is smooth and hence $R^H_{\mu\nu}$ will be too, and since $R^H_{\mu\nu}$ with our reference metric preserves the static isometry, then it follows that $R^H_{\mu\nu}$ must have the behaviour stated above.

In Lorentzian signature, if we are to study only the exterior of the horizon, then the horizon should be regarded as a physical boundary. There is no analog chart to the Riemannian case where the boundary can be smoothly removed without introducing the black hole interior. However, since the Harmonic Einstein equation and the solutions for the metric components are invariant under continuation $t \rightarrow \tau = i t$, then precisely the same boundary conditions for regularity apply in the Lorentzian case \eqref{eq:staticLor} as in the Euclidean case in the static adapted chart \eqref{eq:staticEuc}. Hence we may work directly in Lorentzian signature, where the equations are the same and so are elliptic, and then provide the same boundary conditions there for the metric components at the horizon, taking,
\begin{eqnarray}
\label{eq:staticpolarLor}
ds^2 =  - r^2 V dt^2 + U \left( dr + r \, U_a dx^a \right)^2 + h_{ab} dx^a dx^b
\end{eqnarray}
where $r = 0$ at the horizon, $V > 0 $ outside the horizon and vanishes on it. Then at $r =0$ we again require
 $V, U, U_a, h_{ab}$ are smooth in $r^2$ and $x^a$, and that $V = \kappa^2 U$. Now the constant $\kappa$ precisely gives the surface gravity of the Killing horizon with respect to $\partial / \partial t$. We may manifest the regularity of this horizon in a similar manner to the Euclidean case by performing a change of coordinates,
 \begin{eqnarray}
a = r \sinh{\kappa t} \, , \quad b = r \cosh{\kappa t} 
\end{eqnarray}
giving a chart with coordinates $a,b,x^a$ that now covers the $t = 0$ slice of the Killing horizon, and whose metric components are smooth functions. The essential difference with the Euclidean case is that if we are interested in the exterior of the horizon, then the horizon remains a boundary in this good chart. 
 
Due to the invariance of the components of $R^H_{\mu\nu}$ under $t \rightarrow \tau = i t$,  the tensor $R^H_{\mu\nu}$ shares the same regularity properties as the metric in the Lorentzian context too so that,
 \begin{eqnarray}
R^H =  - r^2 f dt^2 + g \left( dr + r \, g_a dx^a \right)^2 + r_{ab} dx^a dx^b
\end{eqnarray}
 near the chart boundary at $r = 0$ where again $f, g, g_a$ and $r_{ab}$ are smooth in $r^2, x^a$, and in addition $f = \kappa^2 g$. In the Riemannian picture Ricci flow and the Newton method preserve smoothness and lack of boundary at the horizon. We see that equivalently in the Lorentzian picture we have the very nice property that Ricci flow and the Newton method will preserve the surface gravity of the solution.
 
We conclude this discussion with some comments. One attractive feature of the Riemannian approach to static black holes where one removes the horizon boundary by taking periodic time is that there is no boundary associated to the horizon, and the boundary conditions imposed far from the horizon may be taken to fix the size of the time circle. Since the size of the time circle is interpreted as inverse temperature, we see that viewing the system as a boundary value problem we are naturally led to fix physical data. A very nice consequence  of our static Lorentzian discussion is that although we must now view the horizon as a boundary, we are again naturally lead to impose physical data there, namely the surface gravity with respect to $\partial / \partial t$. Asymptotically or on a boundary away from the horizon we impose conditions to fix the value of the function $N$ in \eqref{eq:staticLor}, and this then determines the normalization of $\partial / \partial t$. Thus together these fix the physical data specifying the black hole, and moreover this data is preserved by the Ricci flow and Newton method.

Since in practice (for example \cite{KitchenHeadrickTW}) we choose to use adapted coordinates that manifest the static symmetry even when thinking about the Euclidean formulation of the problem, and then apply boundary conditions where the coordinates degenerate, one might think it makes no difference which signature we think about the static problem in. The mechanics of solving it will be identical in both. Whilst true, there is one important advantage to thinking about the static problem from the Lorentzian perspective, namely that one can consider multiple Killing horizons with respect to $\partial / \partial t$ with \emph{different} surface gravities, each of which is individually preserved by Ricci flow or Newton's method. Formally we could not previously consider this in the Riemannian case, as only one horizon boundary can be removed and made smooth by a choice of periodic time, the remaining horizons becoming conical singularities. By employing the boundary conditions above in the static adapted coordinates of course we may now treat these conical singularities in the Euclidean context, but the motivation to consider Euclidean signature is somewhat diminished.

\section{Stationary vacuum spacetimes with globally timelike Killing vector}
\label{sec:stationary}

We now wish to consider using the methods above to find stationary vacuum solutions. However, now the spacetimes must be considered as Lorentzian from the outset as there is generally no real Euclidean section to a solution.  In this section we begin by considering the case of stationary spacetimes with globally timelike Killing vector (i.e. no horizons or ergo-regions may exist) and will argue that the Harmonic Einstein equation is elliptic. Of course we are ultimately interested in black hole spacetimes which violate such a condition and in the following section \S\ref{sec:blackholes} we consider more general stationary spacetimes which allow horizons and ergo-regions.

Consider the most general stationary metric with Killing vector $T = \partial / \partial t$, which we may write using coordinates adapted to the stationary isometry as,
\begin{eqnarray}
\label{eq:stationarygeneral}
g = - N(x) \left( dt + A_i(x) dx^i \right)^2 + h_{ij}(x) dx^i dx^j
\end{eqnarray}
Now under our assumption that $T$ is globally timelike we have $N > 0$ and we further assume that the function $N$ is bounded. Physically this implies our spacetime has no Killing horizons, and also no ergo-region.
Since $\det{ g_{\mu\nu}} = - N \det h_{ij}$ we see that provided the metric $g$ is Lorentzian and smooth, so that $\det g_{\mu\nu} < 0$ and bounded,  this implies that $\det{h_{ij}} > 0$. We may then regard this metric as a smooth fibration of time over a base manifold $\mathcal{M}$ so that $(\mathcal{M},h)$ is a smooth Riemannian manifold with Euclidean signature metric $h_{ij}$. 

Whilst one might imagine that for a stationary spacetime the natural way to think about the metric is using the ADM ansatz, it is worth noting that the metric above is not of this form, but rather should be thought of as a Kaluza-Klein ansatz over time.\footnote{Interestingly in the seemingly unrelated context of effective field theory used in the context of GR the utility of such a Kaluza-Klein ansatz has been emphasized over the ADM one \cite{KolCLEFT}.} 
Hence the base manifold $(\mathcal{M},h)$ is \emph{not} the submanifold obtained by taking a constant time $t$ slice of the spacetime. Rather it is the geometry one obtains by performing a Kaluza-Klein reduction in the time direction.
As a result of this the second order derivative terms acting on the metric components $g_{\mu\nu}$ in the stationary Harmonic Einstein equation then go as,
\begin{eqnarray}
{R}^H_{\mu\nu} \sim -\frac{1}{2} g^{\alpha\beta} \partial_ \alpha \partial_\beta {g}_{\mu\nu} + \ldots = -\frac{1}{2} h^{ij} \partial_i \partial_j {g}_{\mu\nu} + \ldots
\end{eqnarray}
and we see whilst the metric $g_{\mu\nu}$ is indeed Lorenztian, since there is no dependence on the coordinate $t$, it is actually the metric $h_{ij}$ that controls the character. This immediately implies that the Harmonic Einstein equation $R^H_{\mu\nu} = 0$ is elliptic since $h$ is smooth and Euclidean signature. Thus the stationary problem reduces to an elliptic problem on the Riemannian base manifold $\mathcal{M}$.

However, we also require that $R^H_{\mu\nu}$ is a tensor that is symmetric with respect to the stationary isometry $T$. Without this, Ricci-DeTurck flow and Newton's method will not consistently truncate to the class of stationary metrics \eqref{eq:stationarygeneral}. In order that $R^H_{\mu\nu}$ preserves the symmetry $T$, we choose the reference metric $\bar{g}$ to also be a smooth Lorentzian metric which is stationary with respect to the vector field $T$, so that,
\begin{eqnarray}
\bar{g} = - \bar{N}(x) \left( dt + \bar{A}_i(x) dx^i \right)^2 + \bar{h}_{ij}(x) dx^i dx^j 
\end{eqnarray}
where we also assume here that $T$ is globally timelike and bounded with respect to $\bar{g}$ so that $\bar{N} > 0$ and bounded. Then $\bar{h}_{ij}$ gives a second Riemannian metric on the same manifold $\mathcal{M}$. Since $R^H_{\mu\nu}$ preserves the stationary symmetry, the Ricci-DeTurck flow can be consistently truncated to a parabolic flow on the space of Lorentzian stationary metrics. Since this flow remains diffeomorphic to Ricci flow (subject at least to the normal component of $\xi$ vanishing on any boundaries), we arrive at the interesting result that we may apply Ricci flow to Lorentzian stationary spacetimes. Likewise the Newton method will preserve the symmetry. 

In a situation where the solution we wish to find has a stationary Killing vector that is globally timelike and bounded then nearby to that solution the character of the Harmonic Einstein equation will be elliptic. Subject to imposing suitable boundary conditions on any boundaries or asymptotic regions, one may use the Lorentzian stationary Ricci-DeTurck flow or Newton method to solve for the solution. One must start with an initial guess that has globally timelike bounded stationary Killing field $T$, and then provided that guess is sufficiently good the subsequent Ricci-DeTurck flow or Newton iterations preserve that $T$ is globally timelike. 

We recall that for a solution to the Harmonic Einstein equation to be Ricci flat we require $\xi = 0$. Thus we must ensure that our boundary or asymptotic conditions are compatible with vanishing $\xi$. Let us briefly consider the case of a Dirichlet boundary. The situation is entirely analogous to the static Euclidean case discussed above.
Consider taking coordinates adapted to the boundary, so that,
\begin{eqnarray}
ds^2 = - N(x) \left( dt + A_r(x) dr + A_a(x) dx^a \right)^2 + V dr^2 + V_a dr dx^a + h_{ab}(x) dx^a dx^b
\end{eqnarray}
so that $x^i = ( r, x^a )$ and the boundary is at $r = 0$. Then fixing the induced metric would specify Dirichlet conditions for $N$, $A_a$ and $h_{ab}$. Requiring that $\xi^t$, $\xi^r$ and $\xi^a$ vanish then provides conditions for $V$, $V_a$, and $A_r$.
Thus as in the static case we have precisely fixed both the induced metric and $\xi^\mu = 0$ on the boundary. Following an analogous argument as in \cite{PFLuciettiTW} we expect these boundary conditions are well posed for the elliptic Harmonic Einstein equation.
Provided on other boundaries conditions are set consistent with $\xi$ vanishing we expect solutions where $\xi$ vanishes globally may be found (presuming they exist). Without the maximum principle of the static case \cite{PFLuciettiTW} we cannot rule out solitons with non-vanishing $\xi$, so in a practical context one simply has to test a solution found to see if it is Ricci flat or a soliton.\footnote{
In the static case we have the inequality $\nabla^2 \phi + \xi^\mu \partial_\mu \phi = \nabla_{\mu} \xi_\nu \nabla^\mu \xi^\nu > 0$ where $\phi = | \xi |^2$. However in the stationary case this inequality does not appear to hold as in Lorentzian signature with only stationary symmetry $\nabla_{\mu} \xi_\nu \nabla^\mu \xi^\nu$ is not of definite sign.
}

For completeness we now explicitly give the stationary Ricci-DeTurck flow equations (which of course give the Harmonic Einstein equations at a fixed point $\partial / \partial \lambda = 0$);
\begin{align*}
\frac{\partial N}{ \partial \lambda}  &= \overbrace{\hat{\nabla}^{i}( \partial_{i} N)}^{{}= h^{km} \partial_{k} \partial_{m} N + \dots} - \frac{1}{N} ( \partial^{i} N)( \partial_{i} N) 
- \frac{N^2}{2} F^{ij} F_{ij}  + \hat{\xi}^{k} \partial_{k} N \\
& + \frac{1}{2 N} \bar{h}_{(-1)}^{km} ( \partial_{m} \bar{N})( \partial_{k} N)  +  \bar{h}_{(-1)}^{km} \bar{N} (A^{i} - \bar{A}^{i}) \bar{F}_{jm} \partial_{k} N + \frac{1}{2} (A_{i} - \bar{A}_{i})^2  \bar{h}_{(-1)}^{km}  ( \partial_{m} N)( \partial_{k} N)  
\end{align*}
\begin{align*}
\frac{\partial A_{i}}{ \partial \lambda}  &= \overbrace{\hat{\nabla}^{k} F_{ik} + \hat{\nabla}_{i}(A_{k} \hat{\xi}^{k})+ \hat{\nabla_{i}}(\hat{\nabla}^{p} A_{p} - \bar{\hat{\nabla}}^{p} \bar{A}_{p})}^{{}= h^{km} \partial_{k} \partial_{m} A_{i} + \dots}- \frac{1}{N} F_{ij} \partial^{j}N + \hat{ \xi^{k}}F_{ki}+ \frac{1}{2N}\bar{h}^{km}_{(-1)}F_{ki} \partial_{m} \bar{N} \\ & - \bar{h}^{km}_{(-1)} \bar{N} (A^{j} - \bar{A}^{j})F_{ik}\bar{F}_{jm} + \frac{1}{2} \bar{h}^{km}_{(-1)}(A_{p} - \bar{A}_{p})^2 F_{ki} \partial_{m} \bar{N}+ \hat{\nabla}_{i}\left( \left(\frac{1}{2N}\bar{h}^{mp}_{(-1)}(A_{m} - \bar{A}_{m})\right. \right. \\ & \left. \left. + \frac{1}{ \bar{N}}(A^{p} - \bar{A}^{p}) + \frac{1}{2} \bar{h}^{kp}_{(-1)}(A_{m} - \bar{A}_{m})^2 (A_{k} - \bar{A}_{k})\right) \partial_{p} \bar{N}\right) + \hat{\nabla_{i}}(\bar{h}^{km}_{(-1)} \bar{N} (A^{j} - \bar{A}^{j})(A_{k} - \bar{A}_{k}) \bar{F}_{jm}) 
\end{align*}
\begin{align*}
\frac{\partial h_{ij}}{ \partial \lambda} &= \overbrace{-2 \hat{R}_{ij} + 2\hat{\nabla}_{(i} \hat{\xi}_{j)}}^{{}= h^{km} \partial_{k} \partial_{m} h_{ij} + \dots} + \frac{1}{2 N^2}(\partial_{i} N)( \partial_{j} N)  - N F_{j}^{k} F_{ki}   \\  & \left(\frac{1}{2}h_{ik} \hat{\nabla}_{j}(\frac{1}{N} \bar{h}^{km}_{(-1)} \partial_{m} \bar{N}) + h_{ik} \hat{\nabla}_{j}(\bar{h}^{km}_{(-1)} \bar{N} \bar{F}_{qm} (A^{q} - \bar{A}^{q})) \right. \\ & \left. + \frac{1}{2} h_{ik} \hat{\nabla}_{j}(\bar{h}^{km}_{(-1)}(A^{p} - \bar{A}^{p})^2 \partial_{m} \bar{N}) + (i \leftrightarrow j) \right)
\end{align*}
where indices are contracted and covariant derivatives $\hat{\nabla}$ are with respect to the base metric $h_{ij}$, and we have defined the antisymmetric tensors $F_{ij} \equiv \partial_i A_j - \partial_j A_i$ and $\bar{F}_{ij} \equiv \partial_i \bar{A}_j - \partial_j \bar{A}_i$. Again $\hat{\xi}^i$ is the DeTurck vector of the base metric defined as before in \eqref{eq:baseDeTurck}.

We reiterate that the assumption the reference metric is also stationary with respect to $T$ is responsible for $R^H_{\mu\nu}$ preserving the stationary symmetry and hence the consistent truncation of the Ricci-DeTurck flow to the class \eqref{eq:stationarygeneral} that we see above. However the further assumption that $\bar{g}$ has globally timelike bounded Killing vector $T$, so that $\bar{N} > 0$ and bounded, ensures that the equations are regular and there are no singular terms arising from vanishing or diverging $\bar{N}$ in the above. We note that the components of the DeTurck tensor $R^H_{\mu\nu}$ may be derived from these DeTurck flow equations using the relations,
\begin{eqnarray*}
\frac{\partial N}{ \partial \lambda} &=& -2 R^H_{tt}   \\  
\frac{\partial A_{i}}{ \partial \lambda} &=& - \frac{2}{N} (R^H_{it} - R^H_{tt} A_{i})   \\
\frac{\partial h_{ij}}{ \partial \lambda} &=& -2(R^H_{ij} + R^H_{tt} A_{i} A_{j} - R_{i t} A_{j} - R_{j t} A_{i}) 
\end{eqnarray*}

\section{Stationary black holes and the Harmonic Einstein equation}
\label{sec:blackholes}

We now proceed to consider the case of Ricci flat non-extremal stationary black holes. In the context of the discussion above now the norm of $T$ will vanish either at the horizon itself, assuming that $T$ is a globally timelike Killing vector (such as for certain Kerr-AdS black holes  \cite{Caldarelli:1999xj}), or outside the horizon at the boundary of the ergo-region. Since we are interested in the exterior of the horizon, in the first case we may treat the system described above for globally timelike  $T$ and now regard the horizon as a boundary of the base manifold $\mathcal{M}$ where suitable boundary conditions are required. 
However in the latter, more general case, outside the horizon but inside the ergo-region we have the norm of $T > 0$ and hence $\det{h_{ij}} < 0$. Now the base manifold in the previous section fails to be Riemannian and then our argument above that the Harmonic Einstein equation is elliptic fails. 

In order to make progress we must use the Rigidity property of stationary black holes, proved in $D> 4$ by Ishibashi, Hollands and Wald \cite{Hollands:2006rj} and by Moncrief and Isenberg \cite{Moncrief:2008mr} for various asymptotics including asymptotic flatness. Assume there exists a stationary Killing vector $T$. Then the Rigidity theorem states that for a rotating non-extremal Killing horizon with topology $\mathbb{R} \times \Sigma$, for $\Sigma$ compact, there exists a Killing vector $K$ that commutes with $T$, which is normal to the horizon. Furthermore there exist some number $N \ge 1$ of commuting Killing vectors $R_a$, which also commute with $T$ and generate closed orbits with period $2 \pi$ and $K$ may be written in terms of these as,
\begin{eqnarray}
K = T + \Omega^a R_a
\end{eqnarray}
for some constants $\Omega^a$.  
This result is physically significant as it ensures that the rotation of the horizon is generated by an isometry of the spacetime. Were this not the case one would expect gravitational radiation to be emitted from the region near the horizon and this would presumably violate the assumption of stationarity.

Motivated by the Rigidity theorem we assume that our stationary spacetime, with stationary Killing vector $T$, has $N$ additional Killing vectors $R_a$ for $a = 1,\ldots,N$, which commute amongst themselves and with $T$ and generate rotational or translational isometries. In the former case they generate closed orbits which we take to have period $2 \pi$, and may have axes of rotational symmetry where $R_a$ vanishes. In the latter case they generate non-compact orbits.
Let us take there to be a number of disconnected horizon components, $\mathcal{H}_1 , \ldots , \mathcal{H}_k$. Then Rigidity implies that each component is a Killing horizon with Killing vector given by a linear combination of the isometries $T$ and $R_a$ so that, $K_{\mathcal{H}_m} = T + \Omega_{\mathcal{H}_m}^a R_a$ for some constants $\Omega_{\mathcal{H}_m}^a$ (which may be different for each component).

As a consequence of these assumptions we may write the metric adapting coordinates to the isometries, so that using coordinates $y^A = \{ t, y^a \}$,
\begin{eqnarray}
\label{eq:bhansatz}
d{s}^2 = {g}_{\mu\nu} dX^\mu dX^\nu = {G}_{AB}(x) \left( dy^A + {A}^A_i(x) dx^i \right) \left( dy^B +{A}^B_j(x) dx^j \right) +{h}_{ij}(x) dx^i dx^j
\end{eqnarray}
where $T = \partial / \partial t$ and $R_a = \partial / \partial y^a$. In analogy with the stationary case in the previous section we see that the geometry may be thought of as a fibration of the Killing vector directions over a base manifold $\mathcal{M}$ with metric $h_{ij}$. This base manifold $\mathcal{M}$ is the orbit space of the full Lorentzian spacetime with respect to the isometries $T, R_a$. If $R_a$ generates a compact orbit, then since we have chosen to normalise the period to $2\pi$, then the coordinate $y^a$ is periodic with $y^a \sim y^a + 2 \pi$. Again we note that the metric $h_{ij}$ is not that induced on a constant $y^A$ submanifold of the full spacetime, but rather is the metric one would obtain by performing a `Kaluza-Klein' reduction over these Killing directions.

The full spacetime is Lorentzian and so exterior to the horizons $\det g_{\mu\nu} = \det G_{AB} \det h_{ij} < 0$. 
On physical boundaries or asymptotic regions $T$ is timelike, $R_a$ are spacelike and hence the fiber metric $G_{AB}$ is Lorentzian there, and consequently the base metric is Euclidean. At a horizon $\mathcal{H}_m$, since the norm of $K_{\mathcal{H}_m} = T + \Omega_{\mathcal{H}_m}^a R_a$ vanishes, then $\det G_{AB} = 0$. In addition at axes of symmetry associated to the fixed action of a compact $R_a$, $\det G_{AB}$ will vanish. 

Following the uniqueness theorem treatment of 4$D$ stationary black holes as an elliptic problem on the two dimensional Riemannian orbit space bounded by the horizon and axes of symmetry (as for example discussed in \cite{Carter}) and its generalization to $D$ dimensional metrics with $(D-2)$ commuting Killing vectors \cite{Morisawa:2004tc,Hollands:2007aj} which is treated in the same manner, we make the follow key assumption;
\begin{itemize}
\item We assume that the orbit space base manifold $(\mathcal{M},h)$ is a smooth Riemannian manifold with boundaries given by the horizons and axes of symmetry of the  $R_a$ that generate rotational isometries.
\end{itemize}
A consequence of $\det h_{ij} > 0$ everywhere on $\mathcal{M}$ (including the horizon and axis boundaries), is that $\det G_{AB} \ge 0$ everywhere on $\mathcal{M}$ with it vanishing only at the horizon or axis boundaries of $\mathcal{M}$. 
We note that $\det g_{\mu\nu} = 0$ at the horizons and axes, as one would expect since the chart \eqref{eq:bhansatz} breaks down there.
As Harmark has discussed \cite{Harmark:2009dh}, the structure of $\mathcal{M}$ together with the data $\Omega_{\mathcal{H}_m}^a$ at the horizon boundaries, and the data of which $R_a$ vanishes at the axis boundaries defines a `rod structure' for stationary spacetime and has been conjectured to classify higher dimensional black holes.

\subsection{Ellipticity}

We note that we have not considered the stationary Killing field $T$ to be timelike. In the presence of horizons it will become null on the horizon or be spacelike if the horizon is surrounded by an ergo-region. We reiterate that in the previous section \S\ref{sec:stationary} it was precisely where $T$ failed to be timelike that ellipticity would break down, since the base metric would fail to be Riemannian. The crucial observation is that for our class of stationary spacetimes \eqref{eq:bhansatz},
\begin{eqnarray}
R^H_{AB} & = & -\frac{1}{2} g^{\alpha\beta} \partial_ \alpha \partial_\beta {g}_{AB} + \ldots = -\frac{1}{2} h^{mn} \partial_m\partial_n {G}_{AB} + \ldots \nonumber \\
R^H_{ij} & = & -\frac{1}{2} g^{\alpha\beta} \partial_ \alpha \partial_\beta {g}_{\mu\nu} + \ldots = -\frac{1}{2} h^{mn} \partial_m \partial_n {h}_{ij} + \ldots 
\end{eqnarray}
where again the $\ldots$ represent lower than second order derivative terms. We see the equations have character determined solely by the metric $h_{ij}$, and by our assumption above that the base $\mathcal{M}$ is Riemannian,  this is indeed elliptic.

In analogy with the previous section \ref{sec:stationary}, in order to ensure that $R^H_{\mu\nu}$ shares the symmetries of $g$  we choose the reference metric $\bar{g}$ so that $T, R_a$ are again Killing with respect to it, and obey precisely the same assumptions as above for $g$. Thus we may write,
\begin{eqnarray}
\label{eq:bhref}
d{s}^2 = \bar{g}_{\mu\nu} dX^\mu dX^\nu = \bar{G}_{AB}(x) \left( dy^A + \bar{A}^A_i(x) dx^i \right) \left( dy^B + \bar{A}^B_j(x) dx^j \right) + \bar{h}_{ij}(x) dx^i dx^j
\end{eqnarray}
and we further assume that $(\mathcal{M},\bar{h})$ is a smooth Riemannian manifold. Then the Ricci-DeTurck flow and Newton's method consistently truncate to the Lorentzian stationary spacetimes of the form \eqref{eq:bhansatz}.

We must impose suitable boundary conditions. In addition to the boundaries that define the asymptotics we must also now treat the boundaries at the horizons and axes of symmetries of the rotational Killing vectors. We will shortly discuss these boundary conditions explicitly.
Using the Ricci-DeTurck flow or Newton method if we start from initial data in our stationary class, then for small flow times or updates we expect to remain in this class. In particular we expect $(\mathcal{M},h)$ to remain a Riemannian manifold. Provided this condition holds for the solution of interest, and our initial guess is sufficiently close to this, then we might hope to reach this solution. 

We will now explicitly give the Ricci-DeTurck flow equations, from which the components of $R^H_{\mu\nu}$ can be deduced using,
\begin{eqnarray}
\frac{\partial G_{AB}}{ \partial \lambda} &=& -2 R^H_{AB}  \nonumber  \\  
\frac{\partial A_{j}^{C}}{ \partial \lambda} &=& -2 G^{AC}(R^H_{jA} - R^H_{AB} A^{B}_{j}) \nonumber \\
\frac{\partial h_{ij}}{ \partial \lambda} &=& -2(R^H_{ij} + R^H_{AB} A_{i}^{A} A_{j}^{B} - R^H_{iA} A^{A}_{j} - R^H_{jA} A^{A}_{i})  
\end{eqnarray}
Contracting indices and taking covariant derivatives $\hat{\nabla}$ with respect to the base metric $h_{ij}$, we find,
 \begin{eqnarray}
\frac{\partial G_{AB}}{ \partial \lambda}  &=& \overbrace{\hat{\nabla}^{i}( \partial_{i} G_{AB})}^{{}= h^{mp} \partial_{m} \partial_{p} G_{AB} + \dots} - G^{CD}( \partial^{i} G_{AD})( \partial_{i} G_{CB}) + \frac{1}{2} \bar{h}_{(-1)}^{km} G^{CD}( \partial_{m} \bar{G}_{CD})( \partial_{k} G_{AB}) \nonumber \\ 
&& -\frac{1}{2}  G_{BE} G_{AF} F^{Eij} F_{ij}^{F} + \hat{\xi}^{k} \partial_{k} G_{AB} +  \bar{h}_{(-1)}^{km} \bar{G}_{CD} (A^{Di} - \bar{A}^{Di}) \bar{F}^{C}_{im} \partial_{k} G_{AB} \nonumber\\ 
&& + \frac{1}{2}  \bar{h}^{km}_{(-1)} (A^{Ci} A^{D}_{i} + \bar{A}^{Ci} \bar{A}^{D}_{i}- 2 A^{Ci} \bar{A}^{D}_{i})( \partial_{m} \bar{G}_{CD})( \partial_{k} G_{AB}) \nonumber 
\end{eqnarray}
\begin{eqnarray}
\frac{\partial A_{i}^{C}}{ \partial \lambda}  &=& \overbrace{- \hat{\nabla}^{k} F_{ik}^{C} + \hat{\nabla}_{i} (A^{C}_{k} \hat{\xi}^{k}) + \hat{\nabla}_{i}(\hat{\nabla}^{p} A^{C}_{p} - \bar{\hat{\nabla}}^{p} \bar{A^{C}_{p}})}^{{}=h^{mp} \partial_{m} \partial_{p} A^{C}_{i} + \dots} - G^{AC} F_{ij}^{B} \partial^{j} G_{AB} + \hat{\xi}^{k} F_{ki} ^{C} \nonumber \\  
&&  + \frac{1}{2} \bar{h}^{km}_{(-1)} G^{DE} F_{ki}^{C} \partial_{m} \bar{G}_{DE} - \bar{h}^{km}_{(-1)} \bar{G}_{DE} (A^{jD} - \bar{A}^{jD}) F_{ik}^{C} \bar{F}_{jm}^{E} \nonumber \\ 
&&  + \frac{1}{2} \bar{h}^{km}_{(-1)} (A^{pD} A^{E}_{p} + \bar{A}^{pD} \bar{A}^{E}_{p}- 2 A^{pD} \bar{A}^{E}_{p}) F_{ki}^{C} \partial_{m} \bar{G}_{DE} \nonumber \\ 
& & +  \hat{\nabla}_{i} \left(  \left(\frac{1}{2}\bar{h}^{mp}_{(-1)} G^{DE} (A^{C}_{m}- \bar{A}^{C}_{m}) + \bar{G}^{CE}(A^{pD}- \bar{A}^{pD}) \right. \right. \nonumber \\  
&&  \left. \left.  + \frac{1}{2} \bar{h}^{kp}_{(-1)} (A^{mD} A^{E}_{m} + \bar{A}^{mD} \bar{A}^{E}_{m} - 2 A^{mD} \bar{A}^{E}_{m})(A^{C}_{k} - \bar{A}^{C}_{k})\right) \partial_{p} \bar{G}_{DE} \right) \nonumber \\ 
& & + \hat{\nabla}_{i}(\bar{h}^{km}_{(-1)} \bar{G}_{DE} (A^{jE} - \bar{A}^{jE})(A^{C}_{k} - \bar{A}^{C}_{k}) \bar{F}_{jm}^{D})  \nonumber 
\end{eqnarray}
\begin{eqnarray}
\frac{\partial h_{ij}}{ \partial \lambda} &=& \overbrace{-2 \hat{R}_{ij} +2 \hat{\nabla}_{(i} \hat{\xi}_{j)}}^{{}= h^{mp} \partial_{m} \partial_{p} h_{ij} + \dots} + \frac{1}{2} G^{AB} G^{CD} ( \partial_{i} G_{CB})( \partial_{j} G_{AD}) - G_{AB} F_{j}^{Ak} F^{B}_{ki} \nonumber \\ 
& & +  \left (\frac{1}{2} h_{ik}  \hat{ \nabla}_{j} (G^{AB} \bar{h}^{km}_{(-1)} \partial_{m} \bar{G}_{AB}) + h_{ik} \hat{ \nabla_{j}}( \bar{h}^{km}_{(-1)} \bar{G}_{AB} \bar{F}^{B}_{qm} (A^{qA} - \bar{A}^{qA})) \right. \nonumber \\ 
& & \left.  + \frac{1}{2} h_{ik} \hat{\nabla}_{j}( \bar{h}^{km}_{(-1)} (A^{pA} A^{B}_{p} + \bar{A}^{pA} \bar{A}^{B}_{p} - 2 A^{pA} \bar{A}^{B}_{p}) \partial_{m} \bar{G}_{AB}) + (i \leftrightarrow j ) \right)
\end{eqnarray}
where as before the base DeTurck vector field $\hat{\xi}^i$ is defined as in \eqref{eq:baseDeTurck} and we analogously define $F^A_{ij} \equiv \partial_i A^A_j - \partial_j A^A_i$ and similarly, $\bar{F}^A_{ij} \equiv \partial_i \bar{A}^A_j - \partial_j \bar{A}^A_i$. In the appendix \ref{app:equations} to this paper we present useful intermediate results that lead to these expressions.

\subsection{Reduced stationary case}

We now make a simple observation, namely that if we require our stationary metric to have invariance under the discrete symmetry,
\begin{eqnarray}
t \rightarrow -t \, , \quad y^a \rightarrow -y^a
\end{eqnarray}
then this allows for a consistent truncation of the Harmonic Einstein equation in the sense that the Ricci-DeTurck tensor $R^H_{\mu\nu}$ is also invariant. This symmetry implies that all the $A^A_i$ vanish, and we explicitly see from the above equations that $R^H_{Ai}$ vanishes, as required for the invariance of $R^H_{\mu\nu}$. Consequently the Ricci-DeTurck flow and Newton methods consistently truncate. We term this the `reduced stationary' case.

In this case the Ricci-DeTurck tensor considerably simplifies, and has non-zero components,
\begin{eqnarray}
R^H_{AB}  &= & - \frac{1}{2} \hat{\nabla}^{i}( \partial_{i} G_{AB}) +\frac{1}{2} G^{CD}( \partial^{i} G_{AD})( \partial_{i} G_{CB}) - \frac{1}{4} \bar{h}^{km} G^{CD}( \partial_{m} \bar{G}_{CD})( \partial_{k} G_{AB}) - \frac{1}{2} \hat{\xi}^{k} \partial_{k} G_{AB} \nonumber \\
R^H_{ij} &= &  \hat{R}_{ij} -  \hat{ \nabla}_{(i} \hat{\xi}_{j)} - \frac{1}{4} G^{AB} G^{CD} ( \partial_{i} G_{CB})( \partial_{j} G_{AD}) -  \frac{1}{2} h_{k(i}  \hat{ \nabla}_{j)} (G^{AB} \bar{h}^{km} \partial_{m} \bar{G}_{AB}) 
\end{eqnarray}
with $\xi^A = 0$. For a Dirichlet boundary one imposes the induced metric fixing $G_{AB}$ and the tangential components of $h_{ij}$. The remaining components of $h_{ij}$ are then determined by requiring $\xi^i = 0$.

We note that in 4$D$ the `circularity' theorem \cite{WaldBook} implies all stationary vacuum solutions may be put in this reduced form, given a condition at a single point which is satisfied for asymptotically flat space. More generally all higher dimensional analytic solutions known to us are of this form.

\subsection{Boundary conditions for the Killing horizons and axes of symmetry}
\label{sec:bc}

We now explicitly give the boundary conditions for the components of our stationary spacetime metric \eqref{eq:bhansatz} and reference metric \eqref{eq:bhref} at the Killing horizons or rotational symmetry axes. 
Recall that for the 4$D$ uniqueness theorems the horizon and symmetry axis play the role of boundaries for the Riemannian orbit space, in a very similar manner to our higher dimensional (and cohomogeneity) case here. Therefore we may regard the results in this section on the metric behaviour at the horizons and axes as generalising the boundary conditions in that context  (see for example \cite{Carter}). They are also consistent with the boundary conditions discussed by Harmark using particular coordinates on the base manifold \cite{Harmark:2009dh}.

This is an analogous problem to deducing the smoothness condition for a spherically symmetric function in spherical polar coordinates at the origin. The function depends only on the radial coordinate $r$, but since it is smooth, meaning in Cartesian coordinates $x^i$ it is a smooth $C^\infty$ function of the $x^i$'s, then as $r^2 = x^i x^i$ it cannot be a smooth function of $r$, but rather is smooth in $r^2$. If we require that the function is only $C^2$ then we deduce that the function simply has the Neumann condition $\partial f / \partial r |_{r=0}= 0$. We may perform a similar analysis for a tensor, the only difference being that now the components transform as one moves between a chart which manifests smoothness but not the symmetry, and a chart which manifests symmetry but not the smoothness. The details of this are straightforward and are given in appendix \ref{app:bc} for a smooth $(0,2)$ tensor at a Killing horizon or rotational symmetry axis.
We shall now apply these results to our spacetime metric and reference metric. 

Let us first assume that there is a single Killing horizon, or multiple horizons with common normal Killing vector $K = T + \Omega^a R_a$.  It is then convenient to change coordinates as,
\begin{eqnarray}
\label{eq:trans}
t \, , \; y^a \quad \rightarrow \quad \tilde{t} = t \, , \quad \tilde{y}^a = y^a - \Omega^a t
\end{eqnarray}
so that $K = \partial / \partial \tilde{t}$ and $R_a = \partial / \partial \tilde{y}^a$. We note that if $R_a$ generates a compact orbit, then the coordinate $\tilde{y}^a$ is periodic with $\tilde{y}^a \sim \tilde{y}^a + 2 \pi$.
Now consider a boundary, either due to the vanishing norm of $K$ or a compact $R_a$. We take base coordinates $x^i = ( r , x^{\tilde{i}} )$ adapted to the boundary so that it lies at $r = 0$, and decompose the base metric as,
\begin{eqnarray}
\label{eq:hdecomp}
h_{ij} dx^i dx^j = N dr^2 + r \, N_{\tilde{i}} d r dx^{\tilde{i}} + h_{\tilde{i}\tilde{j}} dx^{\tilde{i}} dx^{\tilde{j}}
\end{eqnarray}
and likewise for the reference metric where $N \rightarrow \bar{N}$, $N_{\tilde{i}} \to \bar{N}_{\tilde{i}}$ and $h_{\tilde{i}\tilde{j}}\to \bar{h}_{\tilde{i}\tilde{j}}$.\\

\noindent {\bf Horizon:}  For a Killing horizon we write the following metric components as,
\begin{eqnarray}
G_{\tilde{t} A} = - r^2  f_A \, , \quad A^{A}_{r} = r g^A \, , 
\end{eqnarray}
for $A = (\tilde{t}, \tilde{y}^a)$ and then let $X = \left\{ f_A \, , \;  g^A \, , \; G_{\tilde{y}^a \tilde{y}^b}  \, , \; 
  A^{A}_{\tilde{i}}\, , \;  
N \, , \; N_{\tilde{i}} \, , \; h_{\tilde{i}\tilde{j}} \right\}
$  be the set of functions describing our metric. Let $\bar{X}$ be the analogous set describing the reference metric.
Then the results of appendix \ref{app:bc} imply that for the metric and reference metric to be smooth we require the following behaviour; the functions $X$ and $\bar{X}$ must be smooth functions of $r^2$ and $x^{\tilde{i}}$ at $r=0$, and furthermore obey the regularity conditions,
\begin{eqnarray}
\label{eq:horizon}
\left( f_{\tilde{t}} - \kappa^2 N \right) |_{r=0} = 0 \, , \quad  \left( \bar{f}_{\tilde{t}} - \kappa^2 \bar{N} \right) |_{r=0} = 0 
\end{eqnarray}
where $\kappa$ is constant and gives the surface gravity with respect to $T$. We note that since both the metric and reference metric are smooth with respect to the same vector field $K$, which is Killing for both, it is the same constant $\kappa$ that must enter the conditions \eqref{eq:horizon} above for both of them.
\\

\noindent {\bf Axis:}
Consider an axis associated to a vanishing compact $R_a$. Without loss of generality choose this to be $R_N$. Then we choose to write,
\begin{eqnarray}
G_{\tilde{y}^N A} = r^2  f_A \, , \quad A^{A}_{r} = r g^A \, , 
\end{eqnarray}
and let $Y = \left\{ f_A \, , \;  g^A \, , \; G_{\tilde{t}\tilde{t}}  \, , \;  G_{\tilde{t}\tilde{y}^{\tilde{a}}}  \, , \;  G_{\tilde{y}^{\tilde{a}}\tilde{y}^{\tilde{b}}}  \, , \;  A^{A}_{\tilde{i}}\, , \;  
N \, , \; N_{\tilde{i}} \, , \; h_{\tilde{i}\tilde{j}} \right\}
$  be the set of functions describing our metric (where $\tilde{a} = 1,\ldots,N-1$). Let $\bar{Y}$ be the set of functions that analogously describe the reference metric. Appendix \ref{app:bc} implies that for a smooth metric and reference metric we must have that the metric functions $Y$ and $\bar{Y}$ are smooth functions of $r^2$ and $x^{\tilde{i}}$ at $r=0$, and in addition we require,
\begin{eqnarray}
\label{eq:axis}
\left( f_{\tilde{y}^N} - N \right) |_{r=0} = 0 \, , \quad  \left( \bar{f}_{\tilde{y}^N} - \bar{N} \right) |_{r=0} = 0
\end{eqnarray}
Of course we obtain analogous conditions for an axis with respect to a different $R_a$. 

We see that if we have a single Killing vector $K = T + \Omega^a R_a$ normal to all horizons, we may use the coordinates $(\tilde{t}, \tilde{y}^a)$ and obtain rather simple boundary  conditions. Of course one can write these conditions in the original $(t, y^a)$ coordinates. The axis conditions take exactly the same form, with replacements $\tilde{t} \rightarrow t$ and $\tilde{y}^a \rightarrow y^a$. If we have multiple Killing horizons with different normals, then one must deduce the boundary conditions for each taking coordinates  as in \eqref{eq:trans} with $\Omega^a$ appropriate to each horizon. \\

\noindent {\bf A horizon meeting an axis:}
It is straightforward to check that the boundary conditions at the meeting of a horizon with an axis, or two axes, are compatible with each other. Here we will explicitly check this for the metric in the former case, for an axis of $R_N$, noting that one obtains similar results for the other cases and for the reference metric. Take coordinates on the base such that the horizon is at $r = 0$ and the axis at $\theta = 0$. Then we write the base metric as,
\begin{eqnarray}
\label{eq:join1}
h_{ij} dx^i dx^j = N dr^2  + M  d\theta^2 + r \theta A dr d\theta + r \, B_{\tilde{i}} d r dx^{\tilde{i}} + \theta \, C_{\tilde{i}} d \theta dx^{\tilde{i}} + h_{\tilde{i}\tilde{j}} dx^{\tilde{i}} dx^{\tilde{j}}
\end{eqnarray}
where now $\tilde{i} = 1, \ldots, D-3$. Then writing,
\begin{eqnarray}
\label{eq:join2}
G_{\tilde{t} \tilde{t} } &=& - r^2  f \, , \quad G_{\tilde{t} \tilde{y}^{\tilde{a}} } = r^2  f_{\tilde{a}} \, ,  \quad G_{\tilde{y}^N \tilde{y}^N} =  \theta^2  g \, , \quad G_{\tilde{y}^N \tilde{y}^{\tilde{a}}} =  \theta^2  g_{\tilde{a}}  \quad G_{\tilde{t} \tilde{y}^N } = r^2 \theta^2 k  \, , \nonumber \\
A^{A}_{r} &=& r \, p^A \; , \quad A^{A}_{\theta} = r \, q^A  \, , 
\end{eqnarray}
our arguments from appendix \ref{app:bc} applied to the horizon $r=0$ and to the axis $\theta = 0$ then imply that the set of functions characterising the metric found in the equations \eqref{eq:join1} and \eqref{eq:join2} above, $N, M,  \ldots , k, p^A, q^A$, together with the remaining components $G_{\tilde{y}^a\tilde{y}^b}$ and $A_{\tilde{i}}^A$, must all be smooth functions in $r^2$, $\theta^2$ and $x^{\tilde{i}}$ near the meeting point $r = \theta = 0$. 
Furthermore regularity requires,
\begin{eqnarray}
\left( f - \kappa^2 N \right) |_{r=0} = 0 \; , \quad \mathrm{and} \quad \left( g - M \right) |_{\theta=0} = 0
\end{eqnarray}
We see that the conditions from each boundary give rise to a consistent set of behaviours above.
In particular it implies that two boundaries (a horizon and axis, or two axes) meet in the base at right-angles.
\\

A very important point is that having introduced boundary conditions for the metric we must check that the conditions that this implies for the vector $\xi^\mu$ are compatible with ensuring the elliptic problem in equation \eqref{eq:Ov} is well posed with trivial solution.
To investigate this we must consider our choice of reference metric \eqref{eq:bhref}, which also 
 is required to be regular and hence is subject to the same  boundary conditions above for its components. Then one can explicitly check that,
\begin{eqnarray}
\xi^r |_{r = 0} = 0 \, , \quad \partial_r \xi^{\tilde{i}} |_{r = 0} = 0 \, , \quad \partial_r \xi^A |_{r = 0}= 0
\end{eqnarray}
both at a horizon and axis of symmetry, which is indeed consistent with well posedness of the associated linear problem \eqref{eq:Ov} and with a trivial  solution.

One may consider the Ricci-DeTurck flow or Newton method operating on the metric $g$ near a horizon. With the choice of reference metric above that has the same isometry $K$ and is also regular at the boundaries, the Ricci-DeTurck tensor will be symmetric under $K$. Consider the Ricci-DeTurck tensor in Cartesian coordinates. Since the metric and reference metric components will be smooth at the fixed point, then the Ricci-DeTurck tensor will be regular there. Thus in our adapted coordinates it will also obey the same regularity conditions as the metric and reference metric. In particular, Ricci-DeTurck flow and Newton's method will preserve regularity, and therefore we have the very elegant result that they will leave the surface gravity constant. The same is true for an axis of symmetry where again regularity is preserved.

We note that we may view the above smoothness conditions in the weaker sense of requiring the Cartesian form to only be $C^2$. In this case instead of finding our various functions above are smooth in $r^2$ and $x^{\tilde{i}}$, we have instead only that these functions obey Neumann boundary conditions, ie. vanishing gradient. In addition we also have the regularity conditions $f_{\tilde{t}} = \kappa ^2 N$  for horizons, and $f_{\tilde{y}^N} = N$ for axes as before. Common folklore is that solutions to elliptic problems are analytic, and hence one might expect that only imposing such $C^2$ boundary conditions one would certainly find the stronger smooth behaviour at the horizons and axes, and presumably the much stronger analytic behaviour.

As a final comment we note that for these $C^\infty$ and $C^2$ conditions, the elliptic system has Neumann boundary conditions for the various functions above, and in addition to this, also the constraints, $f_{\tilde{t}} = \kappa ^2 N$ for horizons, and $f_{\tilde{y}^N} = N$ for axes. One might be concerned that these latter conditions should not be imposed in addition to the Neumann conditions, as this is `too much data' for an elliptic problem. However, we emphasize here that this `fictitious' boundary should be viewed as a regular singular point of the equations, due to the singular terms arising from the vanishing norm of the Killing vector, and hence a usual counting of boundary conditions does not apply. Instead we reiterate that the regularity conditions will be preserved by the Ricci-DeTurck flow and Newton method, and thus it is better to think of these conditions not as boundary conditions, but rather as a restriction of the problem to the class of regular metrics, and using the Ricci-DeTurck flow and Newton method which act within that class.

\subsection{An example: Kerr}

It is instructive to consider the simple example of the Kerr solution in light of our discussion above. The Kerr solution is in reduced stationary form, so that $A^A_i = 0$. In the conventional Boyer-Lindquist coordinates the Kerr metric takes the form,
\begin{eqnarray}
ds^2 = G_{tt} dt^2 + 2 G_{t\phi} dt d\phi + G_{\phi\phi} d\phi^2 + h_{rr} dr^2 + h_{\theta\theta} d\theta^2
\end{eqnarray}
with fiber metric,
\begin{eqnarray}
&&G_{tt} = - \frac{ \left(  \Delta - a^2 \sin^2 \theta \right) }{ \Sigma }  \, , \quad
G_{\phi\phi} =   \sin^2 \theta  \frac{ \left( ( r^2 + a^2 )^2 - \Delta a^2 \sin^2\theta \right) }{ \Sigma} \, , \nonumber \\
&&G_{t\phi} = - a \sin^2\theta \frac{ \left( r^2 + a^2 - \Delta \right) }{ \Sigma}  
\end{eqnarray}
and base,
\begin{eqnarray}
&& h_{rr} = \frac{\Sigma}{\Delta}  \, , \quad h_{\theta\theta} = \Sigma
\end{eqnarray}
where the functions $\Delta, \Sigma$ are defined as $\Delta = r^2 + a^2 - 2 M r$ and $\Sigma = r^2 + a^2 \cos^2\theta$.
The stationary Killing vector $T = \pd{}{t}$ and the rotational Killing vector is $R = \pd{}{\phi}$. 

The base manifold $\mathcal{M}$ has coordinates $r, \theta$ upon which the metric components depend explicitly.
The outer horizon is a boundary of $\mathcal{M}$ and is located at $\Delta = 0$ where $r \equiv r_h =  M + \sqrt{M^2 - a^2}$, and the remaining boundaries are from the axes of rotation at $\theta = 0,\pi$. One finds,
\begin{eqnarray}
\det{G_{AB}} = -(a^2 + r(r-2M)) \sin^2 \theta
\end{eqnarray}
which vanishes at these boundaries. Everywhere in the exterior of the black hole, $r_h < r$ and $0 < \theta < \pi$ we have that $G_{AB}$ has Lorentzian signature and $h_{ij}$ is Euclidean and smooth. The Killing field $K = T + \Omega R$ is tangent to the horizon and timelike near there, where the angular velocity of the horizon is given as, $\Omega = \frac{a}{a^2+r_h^2}$. 

Whilst the $\theta$ coordinate is a regular coordinate on the base manifold at the rotation axes, the radial $r$ coordinate is not at the horizon (since $\Delta \to 0$ there). We therefore define a new radial coordinate, $\rho$, such that $d \rho = dr / \sqrt{\Delta}$, and $\rho = 0$ at the horizon, giving,
\begin{eqnarray}
r = M + \sqrt{M^2 - a^2} \cosh \rho
\end{eqnarray}
so that the components of the base metric $h_{ij}$ are smooth at the horizon boundary. In particular in these coordinates the determinant of the base metric, 
\begin{eqnarray}
h_{ij} dx^i dx^j = \frac{\Sigma}{\Delta} dr^2 + \Sigma d\theta^2 = \Sigma \left( d\rho^2 + d\theta^2 \right)  \quad \implies \quad \det{h_{ij}} = \Sigma^2 \ge  r_h^2
\end{eqnarray}
and thus we see that since $r_h > 0$ the base is indeed a smooth Riemannian manifold everywhere on and in the exterior of the horizon.
Since we have only one horizon it is convenient to use the coordinates discussed above,
\begin{eqnarray}
\tilde{t} = t \, , \quad \tilde{\phi} = \phi - \Omega t
\end{eqnarray}
and then we may confirm that near the horizon, $\rho = 0$, we have,
\begin{align*}
G_{\tilde{t}\tilde{t}} = -\kappa^2 \left( h_{\rho \rho}|_{\rho=0} \right) \rho^2 + \mathcal{O}( \rho^4) 
\end{align*}
\begin{align*}
G_{\tilde{t}\tilde{\phi}}=  \mathcal{O}(\rho^2) \; , \quad
G_{\tilde{\phi} \tilde{\phi}} = \mathcal{O}(1)
\end{align*}
\begin{align*}
h_{\rho \rho} = h_{ \theta \theta} = ( r_h^2 + a^2 \cos^2 \theta) + \mathcal{O}(\rho^2)
\end{align*}
in accord with our boundary behaviour above, 
where $\kappa$ is the surface gravity of the Kerr solution,
\begin{align}
\kappa^2 = \frac{M^2 - a^2}{4M^2 r_{h}^2} \, .
\end{align}
At the axis of symmetry $\theta = 0$ we have,
\begin{align*}
G_{\tilde{t} \tilde{t}} =\mathcal{O} (1) \; , \quad
G_{\tilde {t} \tilde{\phi}} = \mathcal{O}(\theta^2)
\end{align*}
\begin{align*}
G_{\tilde{\phi} \tilde{\phi}} =  \left( h_{\theta\theta} |_{\theta = 0} \right) \theta^2 + \mathcal{O}(\theta^4) 
\end{align*}
\begin{align*}
h_{\rho \rho} = h_{ \theta \theta} = (a^2 + (M + \sqrt{M^2 - a^2} \cosh \rho)^2)  + \mathcal{O} (\theta^2)
\end{align*}
which again agrees with our calculation of axis boundary behaviour. Likewise the same agreement is seen for the axis $\theta = \pi$.

\section{Example application: 4$D$ rotating black holes in a cavity}
\label{sec:example}

In the above we have set out a numerical framework to find general stationary vacuum black holes. We now use the example of 4$D$ rotating black holes in a cavity 
as a toy example to demonstrate the methods discussed may be applied straightforwardly in practice.

We choose to impose Anderson's boundary conditions at the cavity wall \cite{Anderson1}, where we fix the conformal class of the induced metric, and also the trace of the extrinsic curvature of the boundary.  
A canonical choice of cavity is such that the induced metric on the 3 dimensional boundary is conformal to the product of time with a round 2-sphere, 
\begin{eqnarray}
ds^2_{B} & = & - dt^2 + d \theta^2 + \sin^2{\theta} \, d\phi^2
\end{eqnarray}
where $\phi \sim \phi + 2 \pi$. We must then specify the trace of the extrinsic curvature of the boundary which we choose to be constant. If one takes the Schwarzschild solution and cuts it off at finite radius, then the extrinsic curvature of the boundary that is introduced is constant and positive. Hence in the rotating case we also take this trace of extrinsic curvature to be positive. Let us denote this positive constant $\alpha$. By a global scaling we may choose $\alpha$ to take any positive value. For later convenience we choose $\alpha = \sqrt{2}$.

We consider the spacetime to be in the reduced stationary class, with a single Killing horizon with spherical topology which is rotating in the $\phi$ direction, so that $R = \partial / \partial \phi$ is Killing, and $K = T + \Omega R$ is normal to the horizon.
As discussed above it is convenient to use the coordinates,
\begin{eqnarray}
\tilde{t} = t \, , \quad \tilde{\phi} = \phi - \Omega t
\end{eqnarray}
so that the boundary behaviour at the horizon takes a simple form. Now $K = \partial / \partial \tilde{t}$ and $R = \partial / \partial \tilde{\phi}$. We explicitly write an ansatz for the reduced stationary metric as,
\begin{eqnarray}
\label{eq:ansatz}
ds^2 &=& g_{\mu\nu} dx^\mu dx^\nu = G_{AB} dy^A dy^B + h_{ij} dx^i dx^j \\
&=& \left( \begin{matrix} - r^2 V B & r^2 \sin^2{\theta} \, W B \\ r^2 \sin^2{\theta}  \, W B & \sin^2{\theta}  \, S B \end{matrix} \right)_{AB} dy^A dy^B + \left( \begin{matrix} A + \frac{1}{B} r^2 \sin^2{\theta} F^2 & r \sin{\theta} \, F \\ r \sin{\theta}  \, F & B \end{matrix} \right)_{ij} dx^i dx^j \nonumber
\end{eqnarray}
where $y^A = ( \tilde{t}, \tilde{\phi} )$ and $x^i = ( r, \theta )$ and we take the horizon to be located at $r = 0$ where $K$ is null, and the axis of symmetry to be at $\theta = 0, \pi$ where $R$ vanishes.
Finally we must specify the coordinate position of the cavity, and choose this to be at $r=1$. Having chosen this the base $\mathcal{M}$ is then covered by a single chart with domain $0 \le r \le 1$ and $0 \le \theta \le \pi$. However, we further assume the reflection symmetry $\theta \rightarrow \pi - \theta$, to reduce the domain to $0 \le \theta \le \pi/2$. 

By assumption $h_{ij}$ is a Riemannian metric on our coordinate domain. 
Since $K$ is normal to the horizon, we have $V > 0$ in the neighbourhood of the exterior of the horizon. For the metric to be smooth at the horizon, our boundary conditions imply that near $r = 0$  the metric functions $\{ V, S, W, A, B, F \}$ are smooth in $r^2$ and $\theta$ (or for the metric only being $C^2$ we require Neumann boundary conditions in $r$, and $C^2$ in $\theta$), with the additional requirement that,
\begin{eqnarray}
\label{eq:exbc1}
\left(  V B - \kappa^2 A \right)|_{r = 0} = 0
\end{eqnarray}
which specifies the surface gravity. At the axis $\theta = 0$ we require the functions $\{ V, S, W, A, B, F \}$ are smooth in $\theta^2$ and $r$ (again for a $C^2$ metric we require Neumann conditions in $\theta$ and $C^2$ in $r$), with the additional requirement that,
\begin{eqnarray}
\label{eq:exbc2}
S |_{\theta = 0} = 1
\end{eqnarray}
and likewise for the axis at $\theta = \pi $. The reflection symmetry $\theta \rightarrow \pi - \theta$ imposes that $F$ is odd about $\theta = \pi/2$ and the other metric functions are even there.
The induced metric at the cavity boundary $r=1$ in the $\tilde{t}, \tilde{\phi}$ coordinates is conformal to,
\begin{eqnarray}
\label{eq:boundary}
ds^2_{B} & = & ( - 1 + \Omega^2 \sin^2{\theta} ) d\tilde{t}^2 + 2 \Omega \sin^2{\theta} d \tilde{t} d \tilde{\phi} + \sin^2{\theta} d  \tilde{\phi}^2 + d \theta^2 \, ,
\end{eqnarray}
which implies that at $r = 1$ we have Dirichlet conditions,
\begin{eqnarray}
V = (1 - \Omega^2 \sin^2{\theta} ) \, , \quad W = \Omega \, , \quad S = 1 
\end{eqnarray}
where then the function $B$ gives the conformal factor. 
The projector onto the induced metric on the cavity boundary is $L_{\mu\nu} \equiv g_{\mu\nu} - n_\mu n_\nu$ where,
\begin{eqnarray}
n = \frac{1}{\sqrt{A}} \left( \frac{\partial}{\partial r} - \frac{r \sin{\theta} F}{B} \frac{\partial}{\partial \theta} \right)
\end{eqnarray}
is the outer unit normal to the boundary. The trace of the extrinsic curvature, $\alpha$, is then given as
$\alpha \equiv L^{\mu\nu} \nabla_\mu n_\nu$. The remaining three metric functions $B, A, F$ at $r=1$ have boundary conditions determined from simultaneously requiring $\xi^r = \xi^\theta = 0$ together with imposing the extrinsic curvature has constant trace $\alpha = \sqrt{2}$. These take a somewhat complicated form but in essence are coupled oblique boundary conditions. We emphasize that Anderson has proven that these give a regular elliptic system \cite{Anderson1}, ensuring that they give well posed boundary conditions for the numerical problem. Indeed we have encountered no problems with the cavity boundary in this toy example.

We see that the physical moduli of a black hole with such boundary conditions, the surface gravity $\kappa$ and angular rotation $\Omega$, are directly imposed in these boundary conditions.
We see that for $\Omega \ge 1$ the horizon Killing field $K$ is no longer timelike. This implies that the stationary Killing vector $T$ develops an ergo-region for $\Omega \ge 1$. We note that for $\Omega < 1$ the Killing vector $T$ is timelike near the boundary and horizon, although in principle it might become spacelike for some intermediate region. However in the results we now present we see no evidence of such exotic behaviour.

For the reference metric we must choose the same form as in \eqref{eq:ansatz}, with the same boundary conditions. We make a simple explicit choice,
\begin{eqnarray}
\label{eq:ref}
\bar{ds}^2 = \bar{g}_{\mu\nu} dx^\mu dx^\nu &=&  \left( \begin{matrix} - r^2 \left( 1 - \Omega^2 r^2 \sin^2{\theta} \right) & \Omega  r^2 \sin^2{\theta} \\ \Omega r^2 \sin^2{\theta}  & \sin^2{\theta}  \end{matrix} \right)_{AB} dy^A dy^B + \left( \begin{matrix} \frac{1}{\kappa^2} & 0 \\ 0 & 1\end{matrix} \right)_{ij} dx^i dx^j 
\end{eqnarray}
which we note indeed satisfies the smoothness conditions at the horizon and axis, in addition to the requirements \eqref{eq:exbc1} and \eqref{eq:exbc2} above. The constants $\kappa$ and $\Omega$ entering the expression above give the surface gravity and angular rotation. One may compute that the trace of the extrinsic curvature is equal to $\kappa$ for this reference metric. We note that whilst we will fix the trace of the extrinsic curvature for the metric, $\alpha = \sqrt{2}$, the same quantity for the reference metric is not required to be fixed or equal to $\alpha$.

\subsection{Numerical results}

We have used Newton's method to solve the Harmonic Einstein equations in this example, and have also simulated the Lorentzian stationary Ricci-DeTurck flow. Our aim is not to perform high precision numerics, but rather to check that the method behaves as expected. We use simple second order finite difference to represent the Harmonic Einstein equations and boundary conditions numerically. We have used various resolutions up to $160 \times 80$ in the radial and angular directions respectively, and have checked the convergence of the results is consistent with second order scaling.

Whilst in principle smoothness in $r^2$ at the horizon should be preserved by Ricci flow or Newton's method, numerical accuracy and stability is improved if in addition to taking smooth initial data, one also imposes Neumann conditions explicitly on the metric functions at the horizon. We take analogous conditions in $\theta$ at the axis boundaries. Likewise whilst the conditions \eqref{eq:exbc1} and \eqref{eq:exbc2} in principle are again preserved by Ricci flow or Newton's method, we also impose these explicitly on the metric functions as boundary conditions to improve accuracy.

\begin{figure}[ht]
\centering
\includegraphics[width=14cm]{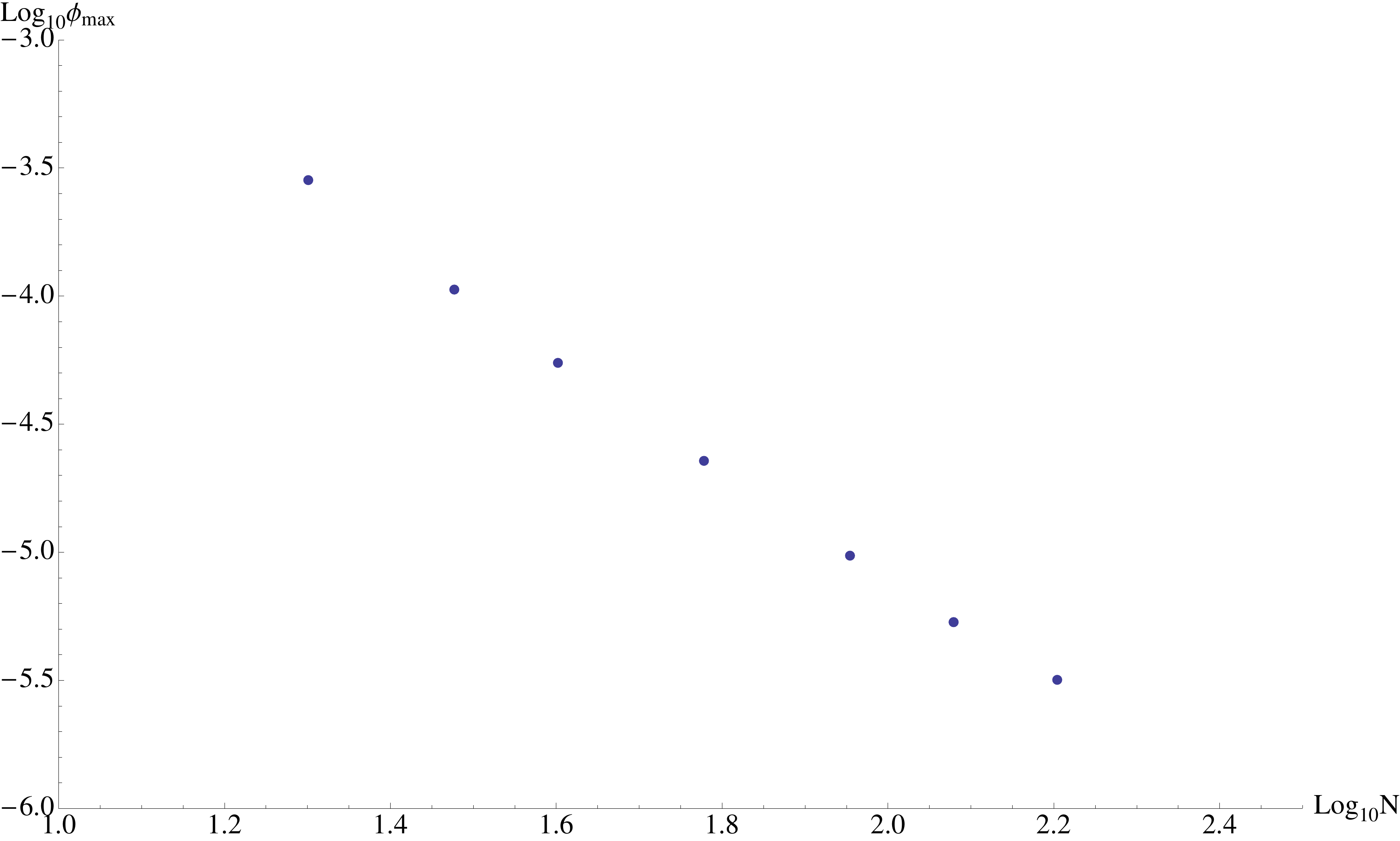} 
\caption{Plot of the log (base 10) of the maximum value of $\phi$ over our domain for a typical solution ($\Omega = 0.6$ and $\kappa^2 = 2$) against log (base 10) of the number of radial lattice points $N$. The number of angular lattice points is $N/2$. The figure shows data for resolutions $20\times10$,  $30\times15$, $40\times20$,  $60\times30$,  $90\times45$, $120\times60$ and $160\times80$. We see that $\phi$, and hence the vector field $\xi$ are consistent with vanishing in the continuum limit, as we would expect for a Ricci flat solution. The same behaviour is seen for all solutions of the Harmonic Einstein equation we have obtained.
}
\label{fig:phi}
\end{figure}

We have found solutions using the Newton method starting with the reference metric as an initial guess. For all these solutions we confirm that the base metric is smooth and Riemannian, although we note that presumably if there was a solution where this was not the case we should not expect to have found it.
We have checked that the solutions found to the Harmonic Einstein equation are indeed Ricci flat, rather than solitons, by ensuring that $\xi^\mu$ is small and consistent with vanishing in the continuum limit. We note that for a reduced stationary metric $\xi^A = 0$, and hence the scalar $\phi = g_{\mu\nu} \xi^\mu \xi^\nu = h_{ij} \xi^i \xi^j > 0$ for $h_{ij}$ being Riemannian. Hence vanishing $\phi$ implies vanishing $\xi^\mu$.
Figure \ref{fig:phi} shows the maximum value of the scalar $\phi$ over our domain for a typical solution, with $\kappa = 2$ and $\Omega = 0.6$ plotted against resolution. We observe similar behaviour for the other values of $\kappa$ and $\Omega$.

\begin{figure}[ht]
\centering
 \includegraphics[width=14cm]{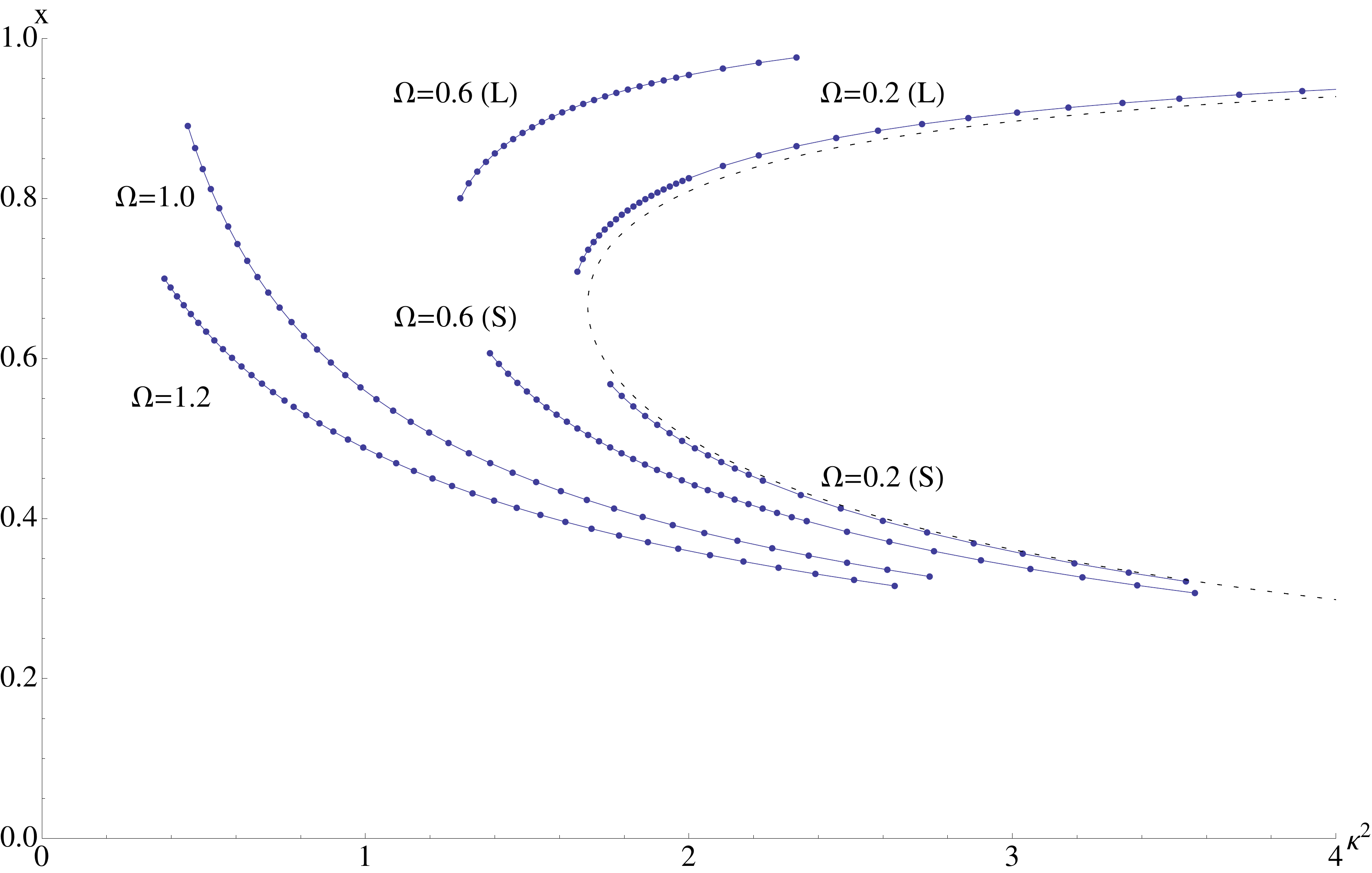} 
\caption{Plot of $x = \sqrt{ \mathcal{A}_h / \mathcal{A}_b }$ against surface gravity $\kappa^2$, where $\mathcal{A}_h$ is the horizon area. Data is shown for numerical solutions with $\Omega = 0.2, 0.6, 1$ and $1.2$. For $\Omega = 0$ the analytic curve is given as a dotted line (the numerical data is omitted as by eye it is indistinguishable from this curve). For $\Omega <1$ there is a minimum value of $\kappa$ which divides the branch of solutions into small (S) and large (L) black holes. The solutions very close to the minimum are difficult to obtain numerically. 
}
\label{fig:plotx}
\end{figure}

\begin{figure}[ht]
\centering
 \includegraphics[width=14cm]{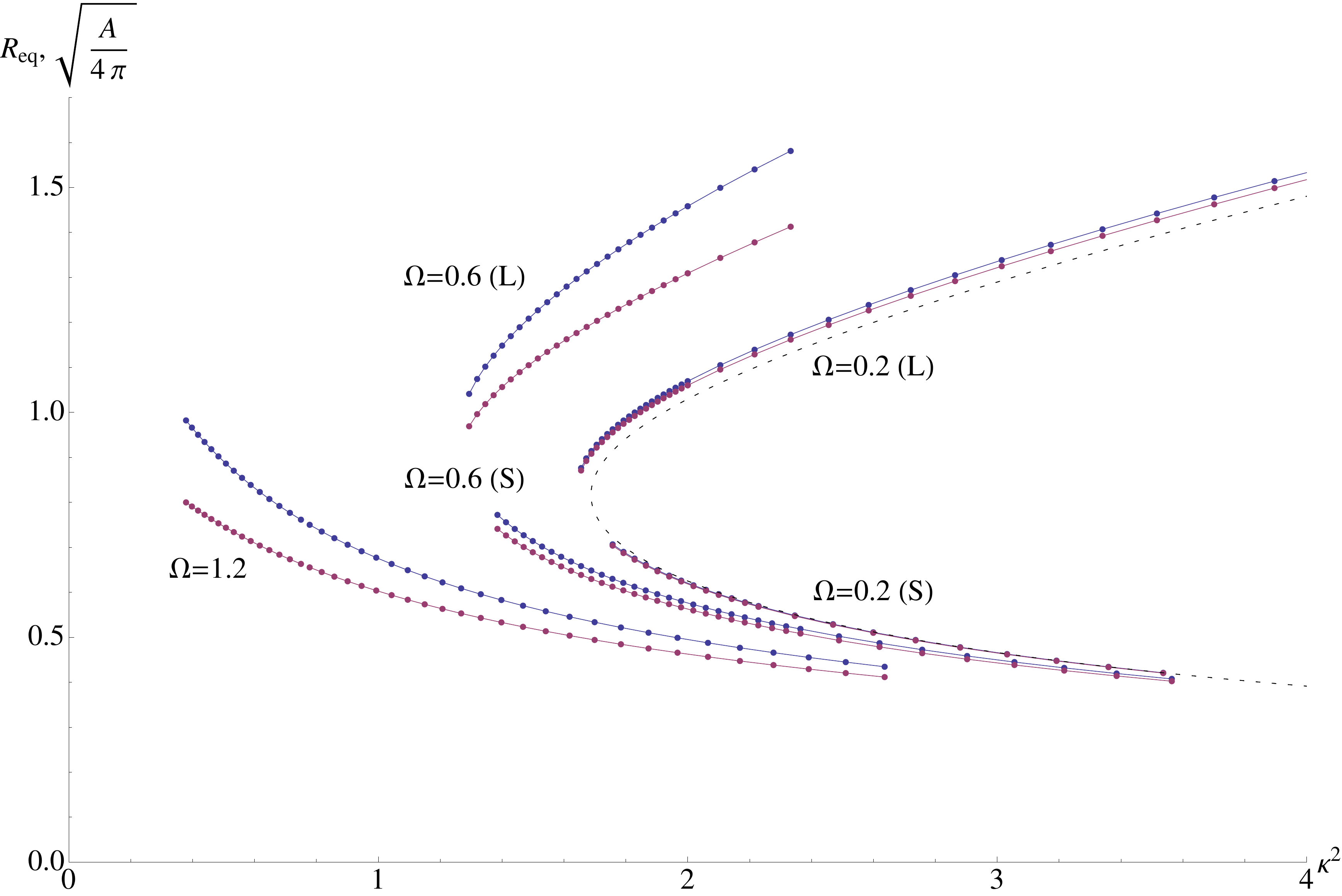} 
\caption{Plots of $R_{eq}$ (red), the equatorial radius, and $\sqrt{\mathcal{A}_h / 4 \pi}$ (blue) against surface gravity $\kappa^2$. Data is shown for $\Omega = 0.2, 0.6$ and $1.2$, with the analytic curve for $\Omega = 0$ drawn as a dotted line. For non zero $\Omega$ and away from the small $x$ limit the horizon becomes deformed from the round sphere to become prolate.}
\label{fig:plotRA}
\end{figure}

Let us define the equatorial radius $R_{eq}$, horizon area $\mathcal{A}_h$, and cavity boundary area $\mathcal{A}_b$,
\begin{eqnarray}
R_{eq} = \sqrt{S B} |_{r = 0 \, , \theta = 0} \, , \quad \mathcal{A}_h = 4\pi \int_{0}^{\pi/2} d\theta \,\left. \sqrt{S} B \sin{\theta} \right|_{r=0} \, , \quad \mathcal{A}_b = 4\pi \int_{0}^{\pi/2} d\theta \,\left. \sqrt{S} B \sin{\theta} \right|_{r=1} \; .
\end{eqnarray}
Let us further define the quantity,
\begin{eqnarray}
x \equiv  \sqrt{ \frac{ \mathcal{A}_h }{ \mathcal{A}_b } } \; ,
\end{eqnarray}
which we may intuitively think of as giving the ratio of the size of the horizon compared to that of the boundary. 

We fix $\Omega$ and then varying $\kappa^2$ to scan through the moduli space of solutions. We have obtained solutions for $\Omega = 0, 0.2, 0.4, 0.6, 0.8, 1.$ and $1.2$, although for graphical clarity we have not presented all this data in the figures that follow.
In figure \eqref{fig:plotx} we plot the quantity $x$ against $\kappa^2$ for fixed $\Omega = 0.2, 0.6, 1$. Similarly in figure \eqref{fig:plotRA} we plot the quantities $R_{eq}$ and $\sqrt{A / 4\pi}$ against $\kappa^2$ for the same solutions.
For $\Omega = 0$ the solutions are static, and therefore with spherical cavity boundary conditions will simply reproduce the Schwarzschild solution. For Schwarzschild one finds,
\begin{eqnarray}
\kappa^2 = \frac{1}{4 x^2 \left( 1 - x \right) } \; , \quad R_{eq} = \frac{1}{2 \alpha} \frac{x \left( 4 - 3 x \right)}{ \sqrt{ 1 - x }} 
\end{eqnarray}
and we observe agreement with the data as we expect. We have not plotted the data points we have computed for $\Omega = 0$ as these lie on the analytic $\Omega = 0$ curves given in the figures.
As we turn on the angular rotation, fixing $\Omega$, we see deviation from this Schwarzschild behaviour, again as we should expect. We note that in the small black hole limit, $x \to 0$, for fixed $\Omega$, we expect to recover the Schwarzschild behaviour and we see this is the case. 

For static solutions, so $\Omega = 0$, there is a minimum surface gravity at $x = 2/3$, and so the solutions divide into the small ($x<2/3$) and large ($x > 2/3$) black holes in an analogous manner to York's construction where one fixes the induced metric on the cavity wall \cite{York:1986it} (rather than the conformal class and $\alpha$). 
The AdS-Kerr solutions of Carter \cite{Carter:1968ks} give an indication of what to expect in the rotating case, where in the usual manner we think of AdS as being a box.
Fixing the AdS length $\ell = 1$, then for $\Omega < 1$ there are small and large solutions and a minimum $\kappa$, and for $\Omega > 1$ there is only one branch of solutions which terminates in an extremal solution \cite{Caldarelli:1999xj}. In addition AdS-Kerr admits a globally timelike Killing vector for $\Omega < 1$, and has an ergo-region for $\Omega > 1$.

We recall that for $\Omega > 1$ we must develop an ergo-region for our solutions. Then in analogy with AdS-Kerr we might expect that for $0 < \Omega < 1$ we have small and large black holes with a minimum surface gravity, whereas for $\Omega > 1$ there is no minimum surface gravity and instead there is an extremal limit where $\kappa \to 0$. This is indeed borne out by our crude numerical results. It would be interesting to confirm with greater accuracy that this transition from minimum surface gravity to extremal limit at fixed $\Omega$ does indeed occur precisely at $\Omega = 1$. Here we have observed that $\Omega = 0.8$ has a minimum surface gravity, and for $\Omega = 1$ we have not found a minimum. We also confirm that for all our solutions with $\Omega < 1$ the vector field $K$ is globally timelike outside the horizon i.e. $G_{\tilde{t}\tilde{t}} < 0$.

We briefly comment on finding the solutions using Newton's method. We begin by finding a solution for some $\kappa$, starting with the reference metric as an initial guess. For $\Omega < 0.6$ starting with $\kappa^2 = 2$ this yields a small black hole solution with boundary trace of extrinsic curvature $\alpha = \sqrt{2}$ as required. We then move along the branch of solutions by perturbing the solution and reference metric to yield a good approximation to a solution with nearby $\kappa$ and the same $\Omega$ and $\alpha$. Using this method we may quickly scan along a branch of solutions to the minimum value of $\kappa$. We could try to find an initial guess in the basin of attraction of a large solution and then scan along this branch. However the approach we take is to extrapolate our small black hole solutions near the minimum $\kappa$ to gain a good guess for a large solution, with $\kappa$ just greater than the minimum. Using this one finds a large solution, and can then scan along the large branch.
The Newton method struggles to find solutions very close to the minimum value of $\kappa$. At the minimum there is a normalisable zero mode of the linearised Harmonic Einstein equation, and near to it there is a low lying mode that renders the linear operator that must be inverted in Newton's method rather ill conditioned. This is why in figures \ref{fig:plotx} and \ref{fig:plotRA} for $\Omega = 0.2$ and $0.6$ the section of the curves connecting the small and large branches are missing.
In order to find black holes with $\Omega > 0.6$ we have found that extrapolating these solutions together with their reference metrics for some fixed $\kappa$, say $\kappa^2 = 2.5$, on the small black hole branch then allows an initial guess for a higher value of $\Omega$ to be found. Having found this new solution branch, one can proceed to iterate this method to find larger $\Omega$ solutions.

\begin{figure}[ht]
\makebox[\textwidth]{
\subfloat[ $\Omega = 0$: Small branch  ]{\includegraphics[width=9cm]{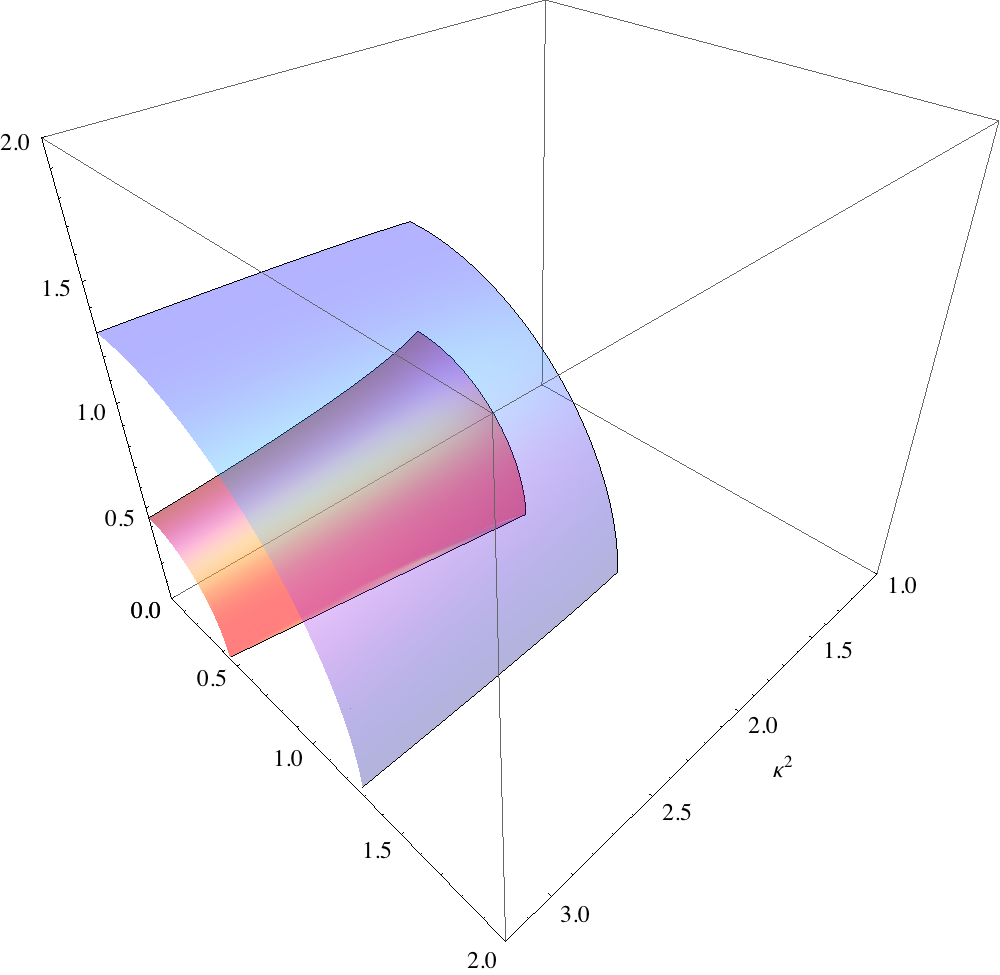}}
\subfloat[ $\Omega = 0$: Large branch ]{\includegraphics[width=9cm]{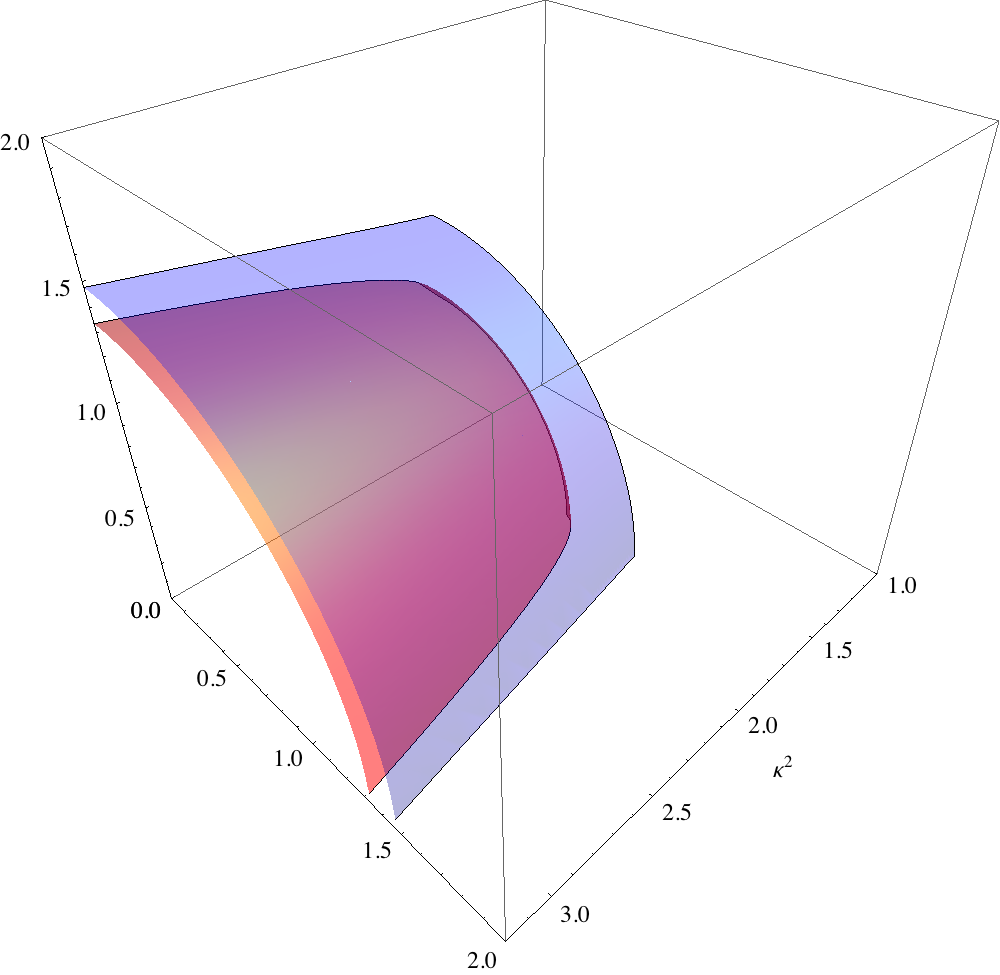}}
 }
 \caption{
Figure showing the upper quadrant of a polar section of the embedding of the horizon (red) and cavity boundary (blue) as a function of $\kappa^2$ for fixed $\Omega = 0$. A slice at constant $\kappa$ should be rotated about the vertical axis, and then reflected in the horizontal plane to obtain the surface of revolution of the embedding in $\mathbb{R}^3$. 
The solutions plotted are for the small black holes (left frame) and the large black holes (right) frame, and being Schwarzschild (as $\Omega = 0$) the horizon and cavity boundary are spherical. This figure should be contrasted with the later figures which give the embeddings for $\Omega > 0$.
 }
\label{fig:embed00}
\end{figure}

\begin{figure}[ht]
\makebox[\textwidth]{
\subfloat[ $\Omega = 0.6$: Small branch  ]{\includegraphics[width=9cm]{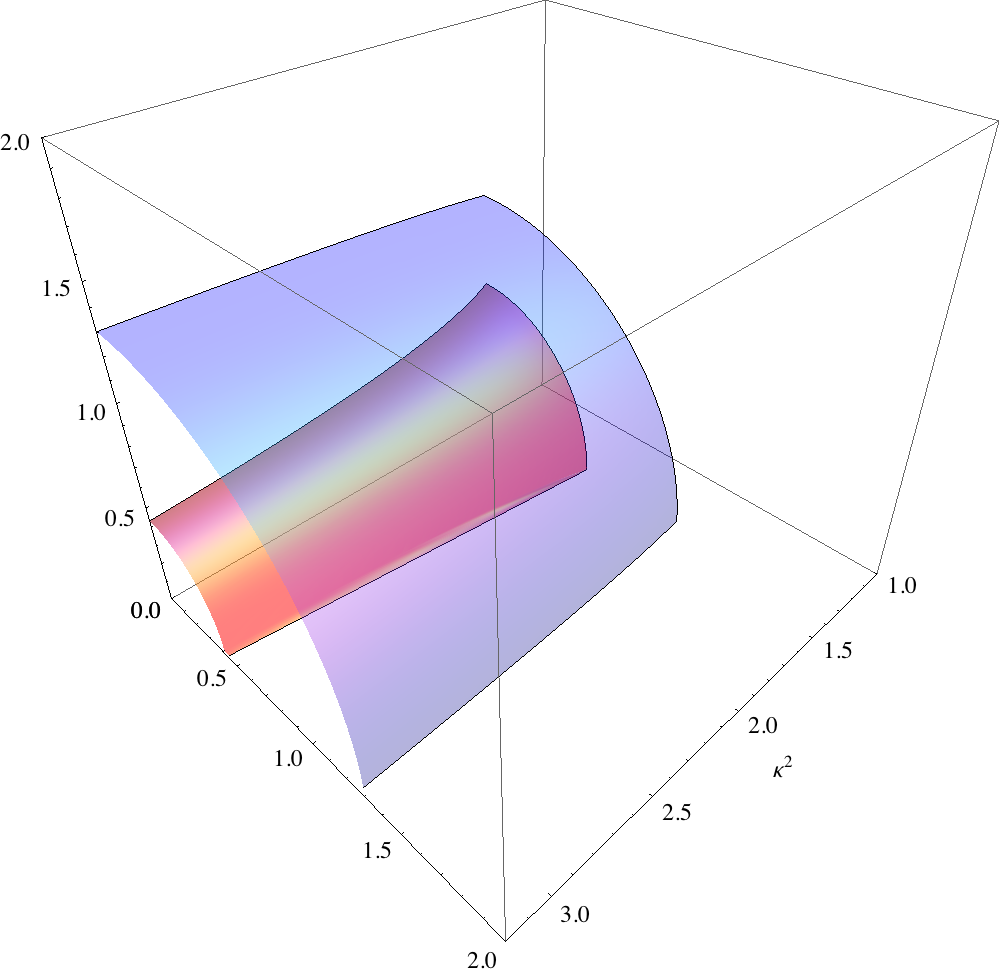}}
\subfloat[ $\Omega = 0.6$: Large branch ]{\includegraphics[width=9cm]{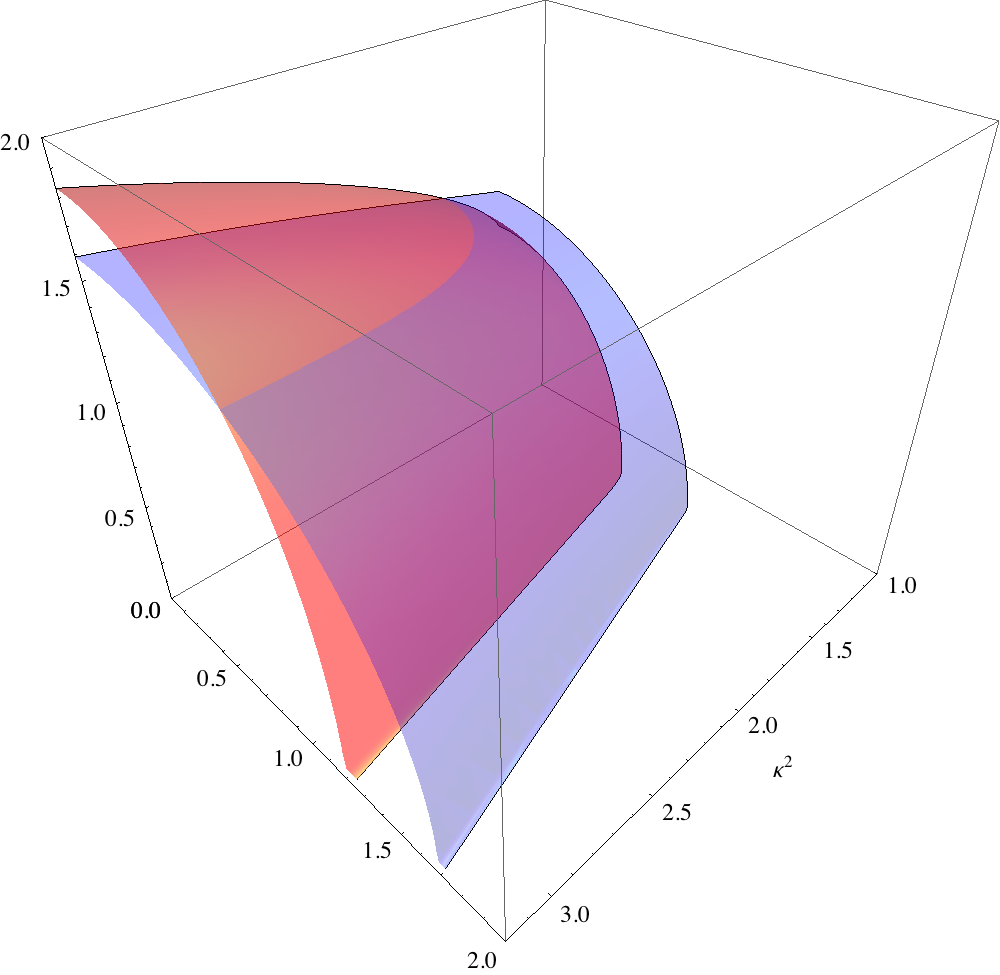}}
 }
\caption{
This figure shows the horizon and cavity boundary embeddings of the small rotating black holes (left frame) and large rotating black holes (right frame) with $\Omega = 0.6$, and should be contrasted with the previous figure depicting Schwarzschild. Interestingly we find for the large black holes the vertical size of the embedding of the horizon may exceed the vertical size of the cavity embedding.
 }
\label{fig:embed06}
\end{figure}

\begin{figure}[ht]
\centering
\includegraphics[width=9cm]{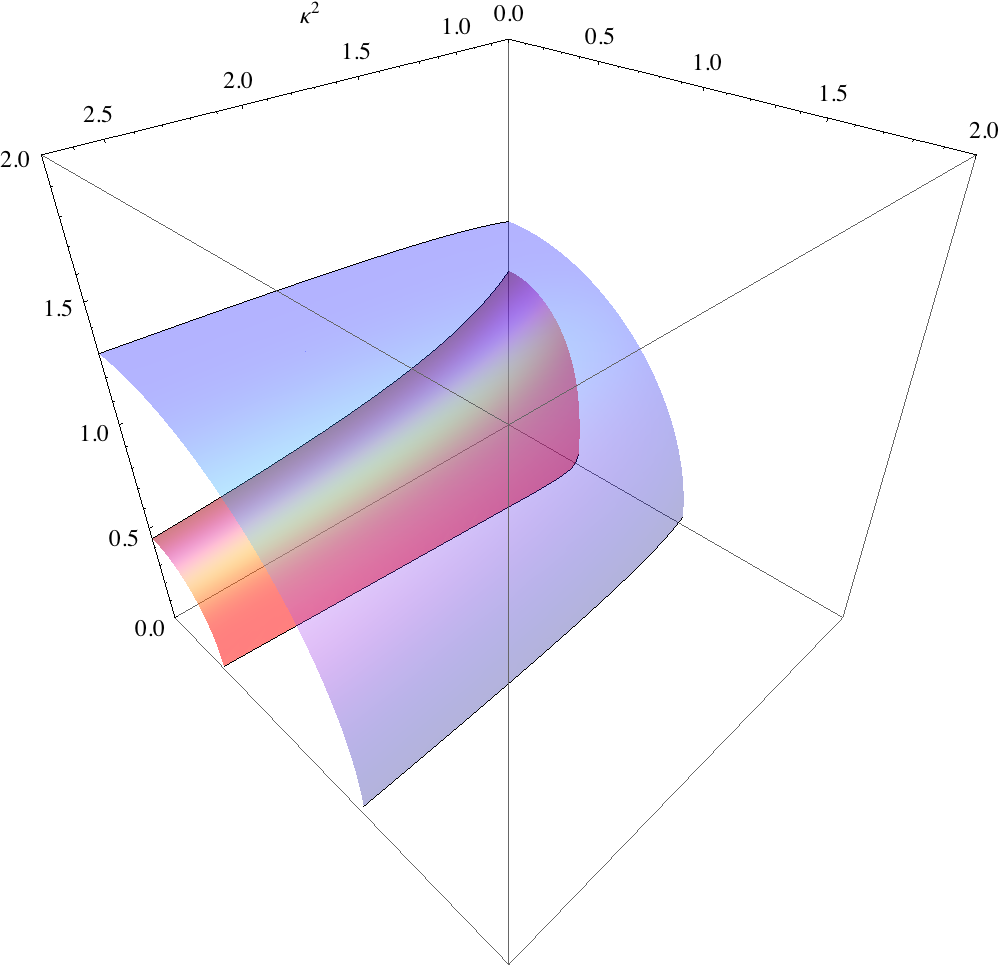} 
\caption{
Figure showing the horizon and cavity boundary embeddings for rotating black holes with $\Omega = 1$. This branch actually extends to lower $\kappa$ although the horizon cannot be globally embedded in $\mathbb{R}^3$ isometrically for $\kappa^2 < 0.7$.
 }
\label{fig:embed10}
\end{figure}

We may isometrically embed the horizon and cavity boundary of these solutions into the Euclidean space $\mathbb{R}^3$ as surfaces of revolution. Taking a polar slice through these embeddings then gives curves in two dimensions representing the geometry of the horizon and boundary. In figures \ref{fig:embed00}, \ref{fig:embed06} and \ref{fig:embed10} we plot the upper quadrant (i.e. $0 \le \theta \le \pi/2$) of these two dimensional embedding curves against $\kappa^2$ for $\Omega = 0, 0.6$ and $1$. Since for $\Omega = 0$ the solutions are simply Schwarzschild, figure \ref{fig:embed00} is included simply for comparison. For the solutions found with $\Omega < 1$ there is both a small and a large branch. Interestingly for $\Omega = 0.6$ we see that for the large solutions with sufficient $\kappa$, the geometry of the horizon is such that its vertical extent in the embedding is actually greater than that of the cavity indicating the geometry of the solution is rather exotic.
For $\Omega = 1$ we find that the horizon may not be embedded isometrically into $\mathbb{R}^3$ for $\kappa^2 < 0.7$ in an analogous manner to the Kerr solution sufficiently near extremality \cite{Gibbons:2009qe}.

We conclude our discussion of these solutions by studying certain stationary Ricci-DeTurck flows. From previous work \cite{HeadrickTW,KitchenHeadrickTW} we know that for static black holes in a spherical cavity with fixed induced metric the small black hole is always unstable having a Euclidean negative mode \cite{GPY,Gregory:2001bd} whereas the large black hole is stable.  
There are two flows for a small black hole generated by this unstable mode. Perturbing the small black hole by the negative mode in one sense generates a flow that asymptotes to the large black hole. By reversing the sign of this perturbation one then generates a second flow that at finite time develops a singularity where the horizon shrinks to zero size. In \cite{HeadrickTW} it was argued that by an appropriate surgery on the manifold one may continue the flow to flat space. 

We note that whilst for fixed induced metric the static spherically symmetric problem (continued to Euclidean signature) is well posed \cite{Holzegel:2007zz}, in our cohomogeneity two example we must take Anderson's boundary conditions to obtain a regular elliptic system. Interestingly we find that for these boundary conditions and  $\Omega < 1$ the large black holes now have one mode of instability for Ricci-DeTurck flow, and the small black holes tested have two. We emphasize that since we are considering the flow in the space of stationary Lorentzian metrics these are \emph{not} Euclidean negative modes, but are rather negative eigenvalue eigenmodes of the Lichnerowicz operator restricted to our class of stationary spacetimes. This implies that to find the large black holes one must tune a one parameter family of initial geometries to reach these unstable fixed points. For the small black holes presumably one must tune two parameters although we have not tried this.

\begin{figure}[ht]
\centering
\makebox[\textwidth]{
\subfloat[$\delta = 2.8$]{\includegraphics[width=8cm]{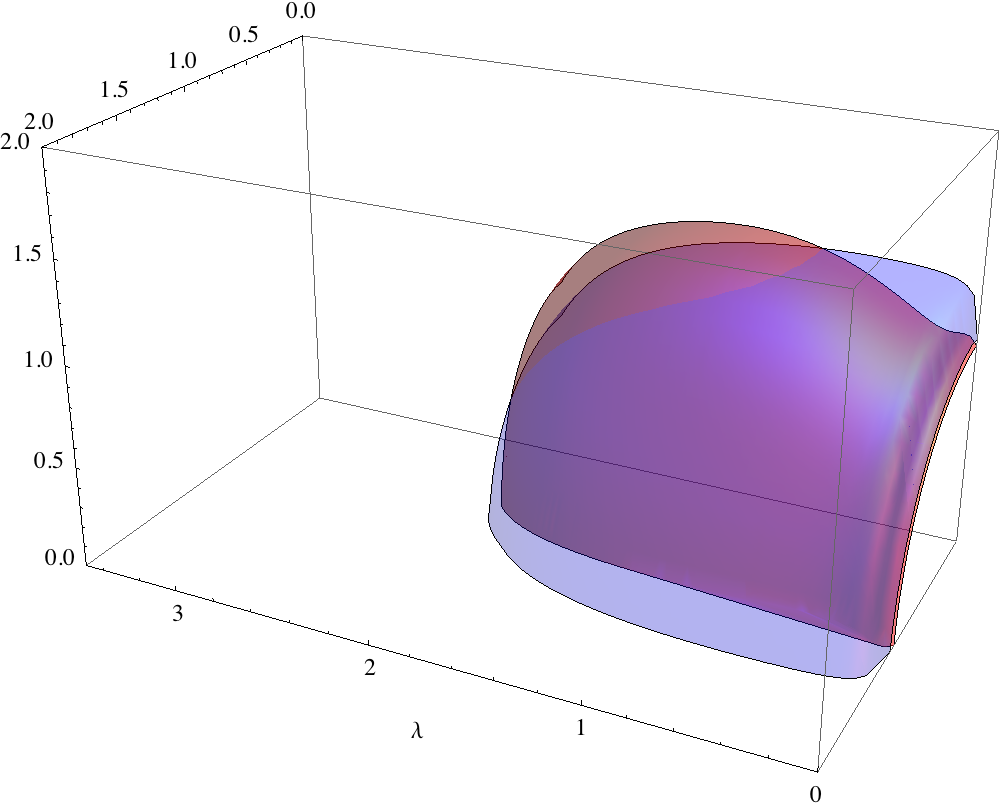}}\subfloat[$\delta = 3.0$]{\includegraphics[width=8cm]{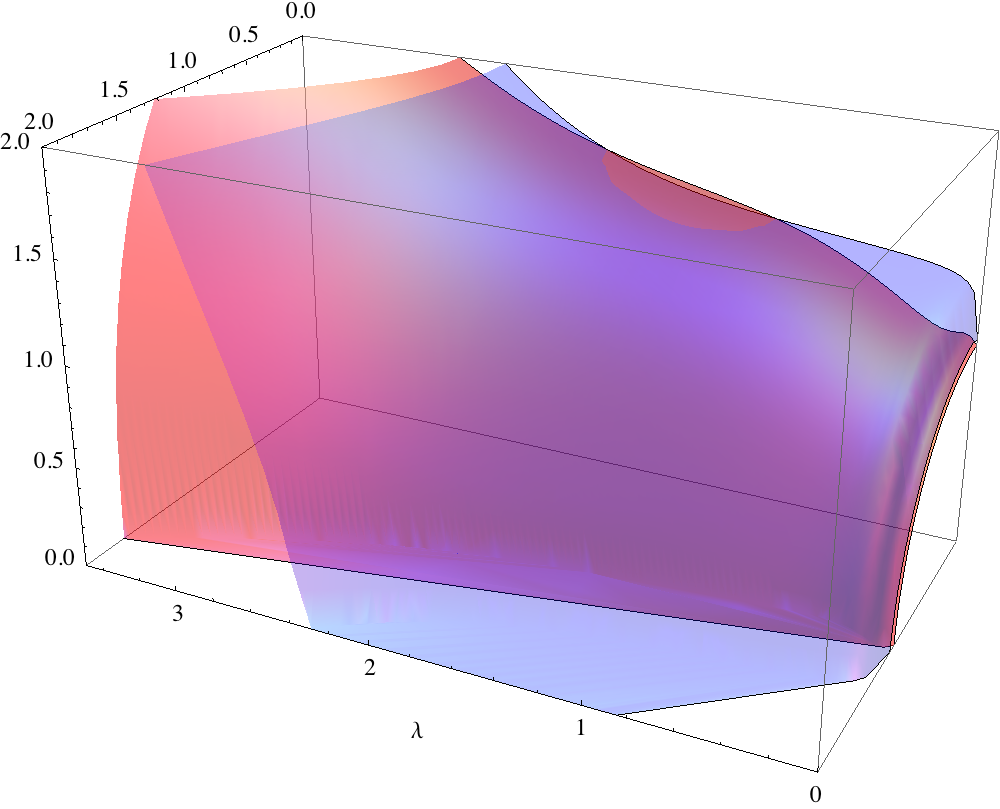}}
}
\makebox[\textwidth]{
\subfloat[$\delta = 2.9$]{\includegraphics[width=8cm]{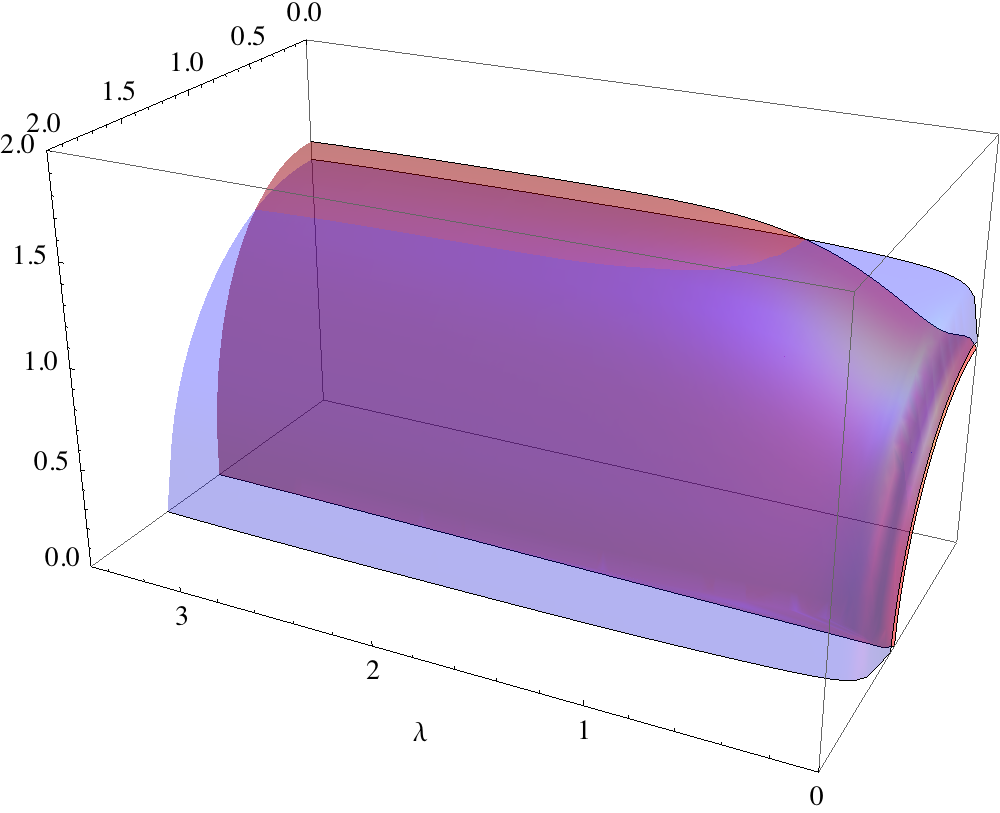}}
}
 \caption{
Figure showing the embeddings of the horizon (red) and cavity boundary (blue) against Ricci flow time $\lambda$ for three values of the parameter $\delta$. For a critical value $\delta \simeq 2.9$ the flow tends to the large rotating black hole solution with $\Omega = 0.6$ and $\kappa^2 = 2$. For smaller $\delta$ the horizon and boundary appear to collapse to a singularity at finite flow time. Conversely for greater $\delta$ they appear to expand in an unbounded manner under Ricci flow. 
 }
\label{fig:flows}
\end{figure}

We now give an example of the Ricci flow algorithm to find a large black hole by tuning a one parameter family of data.
We present results for $\Omega = 0.6$ and $\kappa^2 = 2$, although we expect to see qualitatively similar results for other $\Omega < 1$ and $\kappa$ greater than the minimum surface gravity. Let us take a one parameter family of initial data to be,
\begin{eqnarray}
\label{eq:initdata}
{ds}^2 &=&  \left( \begin{matrix} - r^2 \left( 1 - \Omega^2 r^2 \sin^2{\theta} \right) \left(1 + \delta f \right) & \Omega  r^2 \sin^2{\theta} \\ \Omega r^2 \sin^2{\theta}  & \sin^2{\theta}  \end{matrix} \right)_{AB} dy^A dy^B + \left( \begin{matrix} \frac{1}{\kappa^2} \left(1 + \delta f \right) & 0 \\ 0 & 1 + \delta \left(r f \right)^2\end{matrix} \right)_{ij} dx^i dx^j \nonumber
\end{eqnarray}
with $f = \left(1 - r^2\right)$ and where $\delta$ is the parameter we may tune. We choose $\kappa^2 = 2$ and $\Omega = 0.6$, and then the trace of the extrinsic curvature $\alpha = \sqrt{2}$ as required, independent of the value of $\delta$. 
Likewise the constants $\kappa$ and $\Omega$ entering the expression above give the surface gravity and angular velocity independent of $\delta$. The above metrics also satisfy all our other remaining boundary requirements discussed above.
The reference metric is as before in equation \eqref{eq:ref}. 

We find that we may reach the critical point for $\delta \simeq 2.9$. In figure \ref{fig:flows} we depict embeddings of the horizon and boundary for 3 values of $\delta$, one for slightly less than the critical value ($\delta = 2.8$), one for slightly greater ($\delta = 3.0$), and one close to the critical value ($\delta = 2.9$). We see that for $\delta$ smaller than the critical value the horizon and cavity boundary appear to collapse to a singularity at finite flow time. For $\delta$ greater than the critical value the horizon and boundary appear to expand forever under the Ricci flow. For an appropriately tuned flow we may approach the large rotating black hole solution with $\Omega = 0.6$ and $\kappa^2 = 2$. This is the same solution found by the Newton method. Indeed we see the horizon and cavity embeddings in the tuned flow in this figure match those presented in figure \ref{fig:embed06} for $\kappa^2 = 2$.

\section{Summary}

In this paper we have discussed applying the Harmonic Einstein equation with DeTurck term to the case of Lorentzian static and stationary spacetimes, where Ricci flow or Newton's method acting on these spacetimes can be used as algorithms to solve this equation and yield solutions to the static or stationary vacuum Einstein equations.

Previously static spacetimes have been treated by continuing to periodic Euclidean time, which can remove boundaries associated to non-extremal horizons \cite{KitchenHeadrickTW}. The Harmonic Einstein equation is then solved as an elliptic boundary value problem on a Riemannian manifold, for example, using Ricci flow or Newton's method.

We have firstly shown how to view this procedure for static spacetimes from an entirely Lorentzian point of view. It is straightforward to see, with a suitable choice of reference metric, that the Lorentzian Harmonic Einstein equation consistently truncates to the static case and gives an elliptic system. Whilst the horizon must be treated as a boundary now in Lorentzian signature, by considering the Euclidean continuation, it is easy to deduce Lorentzian boundary conditions for regularity at the horizon. Ricci flow on the space of static spacetimes, or Newton's method can then be applied to solve for static solutions of the vacuum Einstein equations.

Having treated the static case from a manifestly Lorentzian perspective we then considered the stationary case for globally timelike stationary Killing vector. As in the static Lorentzian case it is straightforward to see that for a suitable choice of reference metric, the Harmonic Einstein equation consistently truncates to such stationary spacetimes, again giving an elliptic system. However, we learn that in general, if the stationary Killing field fails to be timelike, such as at horizons or in ergo-regions, it is unclear the system will remain elliptic.

Intuitively we may say that the challenge to ellipticity in the Lorentzian setting is spatial gradients in the direction of motion of the horizon. However for a spacetime to be stationary there can be no such gradients as is formalised in the Rigidity theorems.
 If there were, radiation would be emitted in contradiction to stationarity. Motivated by Rigidity we consider a broad class of stationary spacetimes which are given by a fibration of the orbits of the stationary Killing vector, together with orbits of other mutually commuting Killing vectors that generate rotation or translation and also commute with the stationary field. We then require that the Killing vector normal to a rotating horizon is some linear combination of these vector fields. This is the same class considered by Harmark in order to classify higher dimensional black holes \cite{Harmark:2009dh}. Further motivated by the stationary uniqueness theorems we make the key assumption that the base of this fibration is a Riemannian manifold with boundaries at the horizon and any axes of the commuting rotational symmetries. It is then straightforward to see that for a reference metric of the same form, the Harmonic Einstein equation consistently truncates to this class of stationary spacetimes. We provide the necessary boundary conditions at the Killing horizons or axes of symmetry and demonstrate they are consistent with obtaining vacuum solutions of the Einstein equations rather than only Ricci solitons.
The Ricci-DeTurck flow is then parabolic on this class of Lorentzian stationary spacetimes, and gives an explicit algorithm to solve it, as does Newton's method.

In order to demonstrate that these methods may be applied in practice we have considered a very simple example where analytic solutions are not known, namely a 4$D$ rotating black hole in a spherical cavity. 
We have used Anderson's boundary conditions at the cavity wall, noting that the Dirichlet conditions (i.e. fixed induced metric) often discussed in the static spherically symmetric case are not well posed. 
Solutions were found using Newton's method. Ricci flows of these Lorentzian stationary black hole spacetimes were also performed, and their use demonstrated to construct solutions. We believe these are the first example of Lorentzian Ricci flows that have been studied.

Of course this toy example is only of cohomogeneity two and is 4 dimensional. The main purpose of this paper is precisely to give an elegant geometric framework that applies beyond these cases, and it will be interesting in future work to test these methods in the many physically interesting situations where stationary black holes have been conjectured to exist.

\section*{Acknowledgements}

We would like to thank Pau Figueras and James Lucietti for many illuminating and invaluable discussions.
We would also like to thank Michael Anderson, Barak Kol, Luis Lehner and Eric Woolgar for useful discussions and correspondence.
Furthermore we would like to thank Matthew Headrick for many important discussions and initial collaboration on this work. AA and SK are supported by STFC studentships. TW is supported by an STFC advanced fellowship and Halliday award.

\appendix

\section{Boundary conditions for horizons and axes}
\label{app:bc}

Our elliptic problem is posed on a base $\mathcal{M}$ which has boundaries associated to Killing horizons and axes of symmetry. We must explicitly give boundary conditions that ensure regularity for these. In order to derive the necessary conditions, we perform a similar calculation to that in section \ref{sec:static} where we used the smooth Euclidean continuation of a static black hole to deduce boundary conditions for the Lorentzian problem.

We are interested primarily in regularity of the metric. However we have argued that in the static Lorentzian case, as is clear from the Euclidean picture, that the Ricci-DeTurck tensor also shares the same regularity. Therefore we perform the analysis below for a general $(0,2)$ tensor $J$ which one can take to be the metric or Ricci-DeTurck tensor. 

The procedure is best illustrated by a simple example. Consider a smooth $(0,2)$ tensor, $J$,  which is symmetric with respect to a vector field $R$ that generates $U(1)$ orbits with period $2\pi$ and has fixed action at some point $p$. We will then derive regularity conditions for the components of $J$ in a chart with `polar' coordinates $(r, \alpha)$ adapted to the symmetry so that $R = \pd{}{\alpha}$, and hence the components do not depend on $\alpha$. 

We take $\alpha$ to have period $2 \pi$, and the fixed point of $R$ is $r=0$. To proceed the tensor $J$ is first written in `Cartesian coordinates' $(a,b)$ that do not manifest the $U(1)$ symmetry, but where $J$ has components which are $ \mathbb{C}^{ \infty}$ smooth everywhere, including the fixed point;
\begin{eqnarray}
\label{eq:cart}
J = N(a,b) da^2 + M(a,b) db^2 + K(a,b) da db
\end{eqnarray}
We then introduce the `polar coordinates' $r, \alpha$ that make explicit the $U(1)$ symmetry as,
\begin{eqnarray}
a = r \sin \alpha \, , \;b = r \cos \alpha
\end{eqnarray}
Let us write $J$ in this polar chart as,
\begin{eqnarray}
\label{eq:polar}
ds^2 = r^2 A(r) d\alpha^2 + B(r) dr^2 + r^3 \, C(r) dr \, d\alpha
\end{eqnarray}
where the metric functions $A,B,C$ now depend only on $r$. The symmetry conditions,$ \frac{ \partial A}{ \partial \alpha} = \frac{ \partial B}{ \partial \alpha}=\frac{ \partial C}{ \partial \alpha} =0$ then translate into conditions on the original metric functions $N, M$ and $K$ and evaluation of these gives the regularity conditions, $N= M$ and $K=0$ at the fixed point $a = b = 0$.
When expressed in terms of the polar metric functions these conditions become, $(A - B)|_{r=0} =0$.
Consider the function $A$ which can be given in terms of the Cartesian components as,
\begin{eqnarray}
A = \frac{1}{a^2 + b^2} \left(  a^2 M + b^2 N - a b K \right)
\end{eqnarray}
As a consequence of the above regularity conditions, we see that $A$ is a smooth function of $a,b$. Since it is only a function of $r$, and $r^2 = a^2 + b^2$, it follows that $A$ is a smooth function of $r^2$. Similarly one finds $B, C$ are also smooth in $r^2$. 

Thus we conclude that if we write the tensor $J$ is the polar chart as in \eqref{eq:polar}, we may think of $r = 0$ as a boundary,
where we impose that $A,B,C$ are smooth in $r^2$ there, and $(A - B)|_{r=0} = 0$.

\subsection{Regularity and smoothness at a Killing horizon}

A Killing horizon implies the existence of a normal Killing field $K$, whose isometry group is $\mathbb{R}$, with a fixed point at the bifurcation surface, and whose orbits close on the future and past horizons. 
We consider a smooth $(0,2)$ tensor that is symmetric under $K$, in a chart adapted to the symmetry which covers the exterior of the Killing horizon. The fixed point may be regarded as a boundary of this chart, and we determine regularity conditions on the tensor components there.

We begin in smooth Cartesian coordinates with a $(0,2)$ tensor, $J$, written in components as,
\begin{eqnarray}
\label{eq:Carthoriz}
J  &=&  N da^2 + M db^2 + U da db + Q_{ {i}} da d {x}^{ {i}}  +  R_{ {i}} db d {x}^{ {i}} +  T_{ {i} {j}} d {x}^{ {i}} d {x}^{ {j}}
\end{eqnarray}
where $ {i}=1,...,D-2$. 
We take a Killing horizon with respect to the Killing vector $K$ to be located at $a = b$ and $a = -b$ with bifurcation surface $a = b =0$. Since $J$ is smooth at the horizon these component functions are $C^\infty$ in the neighbourhood of the horizon. The horizon Killing symmetry is not manifest in these coordinates and in analogy with the toy example, we now change to hyperbolic coordinates,
\begin{eqnarray}
a=r \sinh \kappa t \, , \quad  b=r \cosh \kappa t
\end{eqnarray}
so that $K = \partial / \partial t$ and $r = 0$ is the bifurcation surface, with $\kappa$ a constant related to the normalization of $K$ and giving the surface gravity. We write the metric in this polar form as,
\begin{eqnarray}
\label{eq:Polarhoriz}
J &=& - r^2 A d t^2 + B dr^2 +  r^3 C dr d t +  r F_{ {i}} dr d {x}^{ {i}} + r^2 G_{ {i}} d t d {x}^{ {i}} + T_{ {i} {j}} d {x}^{ {i}} d {x}^{ {j}}
\end{eqnarray}
where the component functions are independent of $t$. Repeating the analysis outlined in the toy example one arrives at the conclusion that the functions $A, B, C,  F_{ {i}}, G_{ {i}}, T_{ {i} {j}}$ depend smoothly on $r^2$ and ${x}^{{i}}$, together with the regularity condition,
\begin{eqnarray}
A|_{r=0} &=&  \kappa^2   B|_{r=0}
\end{eqnarray}
Thus we see explicitly that the regularity in the chart \ref{eq:Polarhoriz} depends on the normalization of $K$, and hence the surface gravity.
If we take the tensor $J$ to be the metric, we deduce the regularity conditions on the metric at the Killing horizon. Taking $J$ to be the Ricci-DeTurck tensor we see the behaviour it will exhibit if it shares the symmetry and is regular.

\subsection{Axis of rotation}

We now consider a $(0,2)$ tensor which is symmetric under a Killing field $R$ which generates rotation about an axis with period $2 \pi$. In a chart which manifests the symmetry the axis is fixed under the $U(1)$ action, and may be regarded as a boundary for the chart. We determine the regularity conditions for the components in this chart there. This case is very close to the toy example before. 

We begin with a Cartesian line element of the form
\begin{eqnarray}
\label{eq:Cartaxis}
J  &=&  N da^2 + M db^2 + U da db + Q_{ {i}} da d {x}^{ {i}}  +  R_{ {i}} db d {x}^{ {i}} +  T_{ {i} {j}} d {x}^{ {i}} d {x}^{ {j}}
\end{eqnarray}
and the component functions depend smoothly on $a,b, {x}^{ {i}}$ in the neighbourhood of the axis which we take to be $a = b = 0$. We now change to polar coordinates defined by,
\begin{eqnarray}
a=r \sin \alpha \, , \quad b= r \cos \alpha
\end{eqnarray}
where $ \alpha$ has period $ 2 \pi$ and $R = \partial / \partial \alpha$ and $r = 0$ is the axis. In these coordinates we write the tensor as,
\begin{eqnarray}
\label{eq:Polaraxis}
J &=& r^2 A d \alpha^2 + B dr^2 +  r^3 C dr d \alpha +  r F_{ {i}} dr d {x}^{ {i}} + r^2 G_{ {i}} d \alpha d {x}^{ {i}} + T_{ {i} {j}} d {x}^{ {i}} d {x}^{ {j}}
\end{eqnarray}
and the symmetry is manifest so the metric functions are independent of $\alpha$. Repeating the analysis outlined in the toy example, one finds the metric functions $A, B, C, F_{ {i}}, G_{ {i}}, T_{ {i} {j}}$ are \emph{smooth} functions of $ r^2$ and ${x}^{ {i}}$, together with the regularity condition,
\begin{eqnarray}
A|_{ r=0} &=&  B|_{ r=0}
\end{eqnarray}

\section{Connection Components and Flow Equations}  \label{app:equations}
In this appendix, we give the connection components of the metric \ref{eq:bhansatz} together with the components of the Ricci tensor and $\xi$ vector.
The Christoffel symbols are given by,
\begin{eqnarray}
\Gamma^i_{jk} &=& \hat{ \Gamma}^i_{jk} + \frac{1}{2}h^{im}A_{Aj}F_{km}^{A}+ \frac{1}{2}h^{im}A_{Ak}F_{jm}^{A}- \frac{1}{2}h^{im}A^{A}_{k}A^{C}_{j} \partial_{m} G_{AC} \nonumber \\
\Gamma^i_{AB}&=& - \frac{1}{2} h^{ij}\partial_{j} G_{AB} \nonumber\\
\Gamma^A_{Bi} &=& -\frac{1}{2} A^{Aj} G_{BC} F_{ij}^{C} + \frac{1}{2} A^{Aj}A_{i}^{C} \partial_{j} G_{BC} + \frac{1}{2}G^{AC} \partial_{i} G_{BC} \nonumber\\
\Gamma^A_{ij} &=& -A^{A}_{m} \hat{ \Gamma}^m_{ij} + \frac{1}{2}A^{Ak}A_{Bi}F_{kj}^{B} + \frac{1}{2} A^{Ak}A_{Bj}F_{ki}^{B} + \partial_{(j} A_{i)}^{A} + G^{AB} A_{(i}^{C} \partial_{j)} G_{BC} + \frac{1}{2} A^{Ak} A^{B}_{j} A^{D}_{i} \partial_{k} G_{DB} \nonumber\\
\Gamma^A_{BC}&=& \frac{1}{2} A^{Ai} \partial_{i} G_{BC} \nonumber\\
\Gamma^i_{jA}  &=& - \frac{1}{2}h^{ik} A_{j}^{B} \partial_{k} G_{AB} + \frac{1}{2}h^{ik} G_{AB} F_{jk}^{B} 
\end{eqnarray}
where $ \hat{ \Gamma}^i_{jk}$ is the Christoffel connection of the 'submetric' $h_{ij}$ and $F^{A}_{ij} \equiv \partial_i A_{j}^{A}- \partial_j A_{i}^{A} = \hat{\nabla}_{i} A_{j}^{A} -  \hat{\nabla}_{j} A_{i}^{A} $. The covariant derivative in the latter equation, $ \hat{ \nabla}_{i}$, is defined with respect to the connection $\hat{ \Gamma}^i_{jk}$ of $h_{ij}$ (and is metric compatible with respect to $h_{ij}$). 
Using these results one finds for the decomposition of the Ricci tensor,
\begin{align*}
R_{AB}&= \frac{1}{2} h^{ij} \hat {\nabla}_{i} (\partial_{j} G_{AB})- \frac{1}{4} G^{CD} h^{ip} (\partial_{p} G_{AB})( \partial_{i} G_{CD}) \\ 
& + \frac{1}{2} h^{ij} G^{CD} ( \partial_{j} G_{CB})( \partial_{i} G_{AD})+ \frac{1}{4}h^{mi} h^{jp} G_{BE} G_{AF} F_{mj}^{E} F_{ip}^{F}
\end{align*}
\begin{align*}
R_{iA}-R_{AB} A^{B}_{i}&= \frac{1}{2} h^{jk} G_{AB} \hat{ \nabla}_{j} F_{ik}^{B}+ \frac{1}{2} h^{jk} F_{ij}^{B} \partial_{k} G_{AB} + \frac{1}{4} h^{jm} G^{CD} G_{AB} F_{im}^{B} \partial_{j} G_{CD} 
\end{align*}
\begin{align*}
R_{ij}+R_{AB}A^{A}_{i} A^{B}_{j}- R_{Ai} A^{A}_{j} - R_{Aj} A^{A}_{i} &= \hat{R}_{ij}- \frac{1}{2} G^{CB} \hat{ \nabla}_{j}( \partial_{i} G_{CB}) \\ & 
+ \frac{1}{4} G^{CD} G^{BA} (\partial_{j} G_{DA})( \partial_{i} G_{CB}) + \frac{1}{2} h^{km} G_{AB} F_{jm}^{A} F_{ki}^{B} 
\end{align*}
where $ \hat{R}_{ij}$ is the Ricci tensor computed with respect to $ \hat{ \Gamma}^i_{jk}$. The DeTurck vector $\xi^{ \mu} = g^{ \lambda \nu}( \Gamma^{ \mu}_{ \lambda \nu}- \bar{ \Gamma}^{ \mu}_{ \lambda \nu})$ decomposes as,
\begin{eqnarray}
\xi^{k}&=& \hat{ \xi}^{k} - \frac{1}{2} G^{AB}h^{km} \partial_{m} G_{AB} + \frac{1}{2} G^{AB} \bar{h}^{km}_{(-1)} \partial_{m} \bar{G}_{AB} + h^{ij} \bar{h}^{km}_{(-1)} \bar{G}_{AB} (A_{i}^{A}-\bar{A}_{i}^{A}) \bar{F}_{jm}^{B} \nonumber\\  
& & + \frac{1}{2} h^{ij} \bar{h}^{km}_{(-1)} (A^{A}_{j} A^{B}_{i} + \bar{A}^{A}_{j} \bar{A}^{B}_{i} - 2 A^{A}_{j} \bar{A}^{B}_{i}) \partial_{m} \bar{G}_{AB}  \nonumber\\
\xi^{C}&=& \frac{1}{2}G^{AB} A^{Cj} \partial_{j} G_{AB} - \frac{1}{2}G^{AB} \bar{A}^{Cj} \partial_{j} \bar{G}_{AB}+ h^{ij} ( \hat{\nabla}_{i} A_{j}^{C} - \bar{ \hat{\nabla}}_{i} \bar{A}_{j}^{C}) \nonumber \\ 
& & -  \frac{1}{2} h^{ij} \bar{A}^{Ck} (A^{A}_{j} A^{B}_{i} + \bar{A}^{A}_{j} \bar{A}^{B}_{i} - 2 A^{A}_{j} \bar{A}^{B}_{i}) \partial_{k} \bar{G}_{AB} + h^{ij} \bar{A}^{Ck} \bar{G}_{AB}(A^{A}_{i} - \bar{A}^{A}_{i})  \bar{F}^{B}_{kj}  \nonumber\\ 
&& + h^{ij} \bar{G}^{CB} (A^{A}_{i} - \bar{A}^{A}_{i}) \partial_{j} \bar{G}_{AB} 
\end{eqnarray}
where $ \hat{ \xi}^{k} = h^{ij}( \hat{ \Gamma}^{k}_{ij}- \bar{ \hat{ \Gamma}}^{k}_{ij})$ and as usual, an overbar indicates that the quantity in question is evaluated in the reference metric. 

As discussed in the main text, the flow equations for the various metric components of interest decompose as,
\begin{eqnarray*}
\frac{\partial G_{AB}}{ \partial \lambda} &=& -2 R_{AB} + 2 \nabla_{(A} \xi_{B)}  \\  
\frac{\partial A_{j}^{C}}{ \partial \lambda} &=& -2 G^{AC}(R_{jA} - R_{AB} A^{B}_{j}) + 2 G^{AC}( \nabla_{(A} \xi_{j)} -  \nabla_{(A} \xi_{B)} A^{B}_{j} )  \\
\frac{\partial h_{ij}}{ \partial \lambda} &=& -2(R_{ij} + R_{AB} A_{i}^{A} A_{j}^{B} - R_{iA} A^{A}_{j} - R_{jA} A^{A}_{i})  \\  && + 2 ( \nabla_{(i} \xi_{j)} + \nabla_{(A} \xi_{B)} A_{i}^{A} A_{j}^{B} -  \nabla_{(B} \xi_{i)} A_{j}^{B} - \nabla_{(B} \xi_{j)} A_{i}^{B}  )
\end{eqnarray*}

This form is particularly useful as the linear combinations of the components of $ \nabla_{( \mu} \xi_{ \nu )}$ that arise take a relatively simple form. Explicitly one finds that,
\begin{align*}
2 \nabla_{(A} \xi_{B)} &= \hat{\xi}^{k} \partial_{k} G_{AB} - \frac{1}{2} G^{CD} (\partial^{k}G_{CD}) (\partial_{k} G_{AB}) + \frac{1}{2} G^{CD} \bar{h}^{km}_{(-1)} (\partial_{m} \bar{G}_{CD})(\partial_{k} G_{AB}) \\ & +\bar{h}^{km}_{(-1)} \bar{G}_{CD} \bar{F}^{D}_{jm} (A^{jC} - \bar{A}^{jC}) \partial_{k} G_{AB} \\ & + \frac{1}{2} \bar{h}^{km}_{(-1)} (A^{iC} A_{i}^{D} + \bar{A}^{iC} \bar{A}_{i}^{D} - 2 A^{iC} \bar{A}_{i}^{D}) (\partial_{m} \bar{G}_{CD})( \partial_{k} G_{AB})
\end{align*}
\begin{align*}
2 (\nabla_{(i} \xi_{A)} &-  \nabla_{(A} \xi_{B)} A^{B}_{i})  = G_{AC} \hat{\nabla}_{i}(A_{k}^{C} \hat{\xi}^{k}) + G_{AC} \hat{\nabla}_{i}(\hat{\nabla}^{p} A^{C}_{p} - \bar{\hat{\nabla}}^{p} \bar{A^{C}_{p}}) \\ & + G_{AC} \hat{\nabla}_{i} \left(  \left(\frac{1}{2}\bar{h}^{mp}_{(-1)} G^{DE} (A^{C}_{m}- \bar{A}^{C}_{m}) + \bar{G}^{CE}(A^{pD}- \bar{A}^{pD}) \right. \right. \nonumber \\  
&  \left. \left.  + \frac{1}{2} \bar{h}^{kp}_{(-1)} (A^{mD} A^{E}_{m} + \bar{A}^{mD} \bar{A}^{E}_{m} - 2 A^{mD} \bar{A}^{E}_{m})(A^{C}_{k} - \bar{A}^{C}_{k})\right) \partial_{p} \bar{G}_{DE} \right) \nonumber \\ 
& + G_{AC} \hat{\nabla}_{i}(\bar{h}^{km}_{(-1)} \bar{G}_{DE} (A^{jE} - \bar{A}^{jE})(A^{C}_{k} - \bar{A}^{C}_{k}) \bar{F}_{jm}^{D})  \nonumber \\ &  + G_{AC} \hat{\xi}^{k} F_{ki} ^{C} - \frac{1}{2} G_{AC} G^{DE} F_{ki} ^{C} ( \partial^{k} G_{DE}) \nonumber \\ &  + \frac{1}{2} \bar{h}^{km}_{(-1)}G_{AC} G^{DE} F_{ki}^{C} \partial_{m} \bar{G}_{DE} \nonumber \\ &  - \bar{h}^{km}_{(-1)} G_{AC} \bar{G}_{DE} (A^{jD} - \bar{A}^{jD}) F_{ik}^{C} \bar{F}_{jm}^{E} \nonumber \\ 
&  + \frac{1}{2} \bar{h}^{km}_{(-1)}G_{AC} (A^{pD} A^{E}_{p} + \bar{A}^{pD} \bar{A}^{E}_{p}- 2 A^{pD} \bar{A}^{E}_{p}) F_{ki}^{C} \partial_{m} \bar{G}_{DE}
\end{align*}
\begin{align*}
2 ( \nabla_{(i} \xi_{j)} &+ \nabla_{(A} \xi_{B)} A_{i}^{A} A_{j}^{B}  -  \nabla_{(B} \xi_{i)} A_{j}^{B} - \nabla_{(B} \xi_{j)} A_{i}^{B}  ) = 2 \hat{\nabla}_{(i} \hat{\xi}_{j)}  + G^{AB} G^{CD} ( \partial_{i} G_{CB})( \partial_{j} G_{AD}) \nonumber \\ 
& - G^{AB} \hat{\nabla}_{i} ( \partial_{j} G_{AB}) + \left ( \frac{1}{2} h_{ik}  \hat{ \nabla}_{j} (G^{AB} \bar{h}^{km}_{(-1)} \partial_{m} \bar{G}_{AB}) + h_{ik} \hat{ \nabla}_{j}( \bar{h}^{km}_{(-1)} \bar{G}_{AB} \bar{F}^{B}_{qm} (A^{qA} - \bar{A}^{qA})) \right. \nonumber \\ 
& \left.  + \frac{1}{2} h_{ik} \hat{\nabla}_{j}( \bar{h}^{km}_{(-1)} (A^{pA} A^{B}_{p} + \bar{A}^{pA} \bar{A}^{B}_{p} - 2 A^{pA} \bar{A}^{B}_{p}) \partial_{m} \bar{G}_{AB}) + (i \leftrightarrow j ) \right)
\end{align*}
where we note that in these latter three expressions, all 'A term' base indices have been contracted with the base metric $h_{ij}$ as appropriate. Using these results, one arrives at the flow equations in the main body of the paper, contracted in the same manner.
\bibliographystyle{JHEP}
\bibliography{stationary}


\end{document}